\documentclass[osajnl,twocolumn,showpacs,superscriptaddress,10pt]{revtex4-1} 
\usepackage{amsmath,amssymb,graphicx}
\usepackage{lipsum}

\usepackage{pgf,tikz}								
\usetikzlibrary{arrows,fit,patterns}								
\usetikzlibrary{decorations.pathreplacing}
\usepackage[draft,english]{fixme}
\usepackage[colorlinks=true,citecolor=black,linkcolor=black,urlcolor=blue,pdfborder={0 0 0}]{hyperref}
\usepackage{multirow}
\usepackage[labelsep=space,small,bf,justification=raggedright]{caption}
\usepackage{lmodern} 
\usepackage{microtype} 
\usepackage[flushleft]{threeparttable}

\newcommand{\vekn}[1]{\boldsymbol{\mathrm{#1}}}  	
\newcommand{\vekE}{\vekn{E}} 
\newcommand{\vekH}{\vekn{H}} 
\newcommand{\vekr}{\vekn{r}} 
\newcommand{\ddd}{\, \mathrm{d}} 
\newcommand{\ci}{\mathrm{i}} 
\newcommand{\omegat}{\tilde{\omega}} 

\begin{document}

\title{A roundtrip matrix method for calculating the leaky resonant modes of open nanophotonic structures}

\author{Jakob Rosenkrantz de Lasson}
\email{Corresponding author: jakob@jakobrdl.dk}
\author{Philip Tr\o st Kristensen}
\author{Jesper M\o rk}
\author{Niels Gregersen}
\affiliation{DTU Fotonik, Department of Photonics Engineering, Technical University of Denmark, \O rsteds Plads, Building 343, DK-2800 Kongens Lyngby, Denmark}
\date{\today}

\begin{abstract}
We present a numerical method for calculating quasi-normal modes of open nanophotonic structures. The method is based on scattering matrices and a unity eigenvalue of the roundtrip matrix of an internal cavity, and we develop it in detail with electromagnetic fields expanded on Bloch modes of periodic structures. This procedure is simpler to implement numerically and more intuitive than previous scattering matrix methods, and any routine based on scattering matrices can benefit from the method. We demonstrate the calculation of quasi-normal modes for two-dimensional photonic crystals where cavities are side-coupled and in-line-coupled to an infinite W1 waveguide and show that the scattering spectrum of these types of cavities can be reconstructed from the complex quasi-normal mode frequency.
\end{abstract}


\maketitle 

\section{Introduction}\label{Sec:Introduction}
Resonant modes are central in nanophotonics and quantum optics and pave the way for enhanced light-matter interactions with potential applications in energy efficient photovoltaics, integrated photonic circuits and quantum information technology. Examples of resonant modes include the well-known Mie resonances of spherical objects~\cite{Mie1908,Lai1990} and localized surface plasmons of plasmonic nanoparticles, with applications in photovoltaics~\cite{Atwater2010}, surface-enhanced Raman scattering~\cite{Mirsaleh-Kohan2012} or as ``plasmon rulers''~\cite{Yang2010}. Likewise, the optical modes of microcavities in micropillars or photonic crystals (PhCs) have been used for enhancement of the Purcell effect of quantum emitters~\cite{Gerard1999} and for realizing cavity quantum electrodynamics experiments~\cite{Lermer2012} and single-photon emission~\cite{Dousse2010,Schwagmann2012} as well as for demonstrating nanolasers~\cite{Strauf2011} and optical switching~\cite{Nozaki2010}. Inherent to the resonant modes is their leaky nature; they dissipate energy into heat or by radiation into the environment. The leakiness is typically quantified via the $Q$ factor, which measures the stored energy relative to the energy lost per cycle~\cite{Jackson1998Chap8}, and resonant modes of realistic resonators always exhibit finite $Q$ factors. Throughout the literature, the resonant modes are typically calculated in an indirect way using standard time domain or frequency domain methods; in the time domain, short pulses are used to excite the resonant modes, and in the frequency domain scattering calculations are used to excite the resonant modes at different frequencies. Innate to these procedures is an ad hoc choice of the spatial shape and polarization of the excitation pulse or field to excite the resonant modes, which is well-known to pose problems when degenerate or spectrally close resonant modes exist in the structure under consideration. Also, certain classes of modes may not respond to the chosen excitation; for example, ``dark modes" of plasmonic structures do not respond to excitations from the far-field such as plane waves~\cite{deLasson2013}. Additionally, both time and frequency domain methods suffer from the difficulty of resolving time signals or spectra sufficiently to obtain accurate values of $Q$ for underlying high-$Q$ modes. 

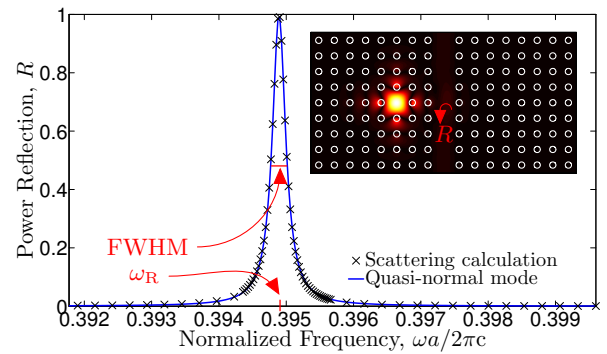
\begin{figure}[htb!]
\centering

\begin{tikzpicture} [line cap=round,line join=round,>=triangle 45,x=1.0cm,y=1.0cm, scale=0.47]
%
%

\node (img) at (0,0){\includegraphics[scale=0.19]{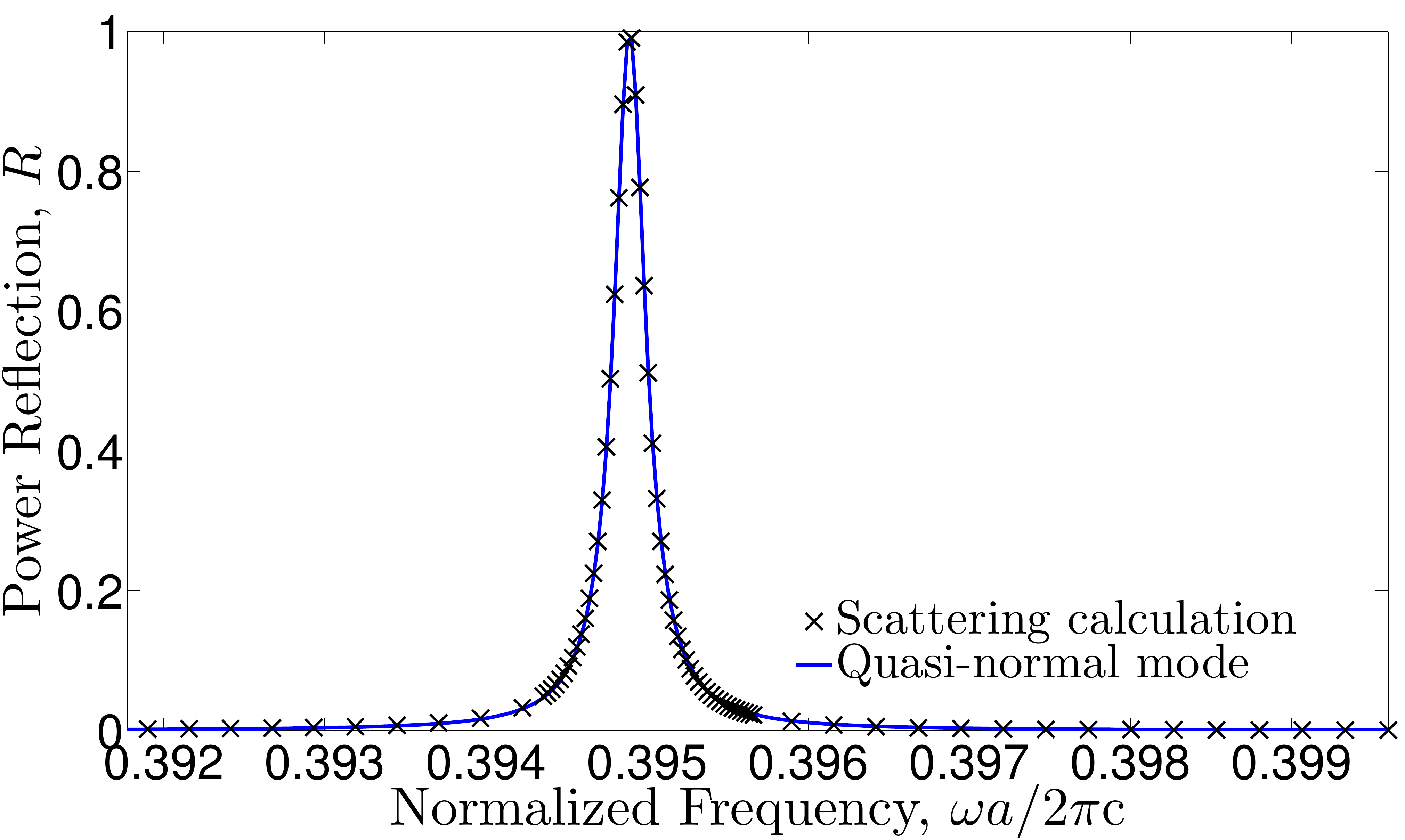}};
\node (img) at (3.8,2.1){\includegraphics[scale=0.09]{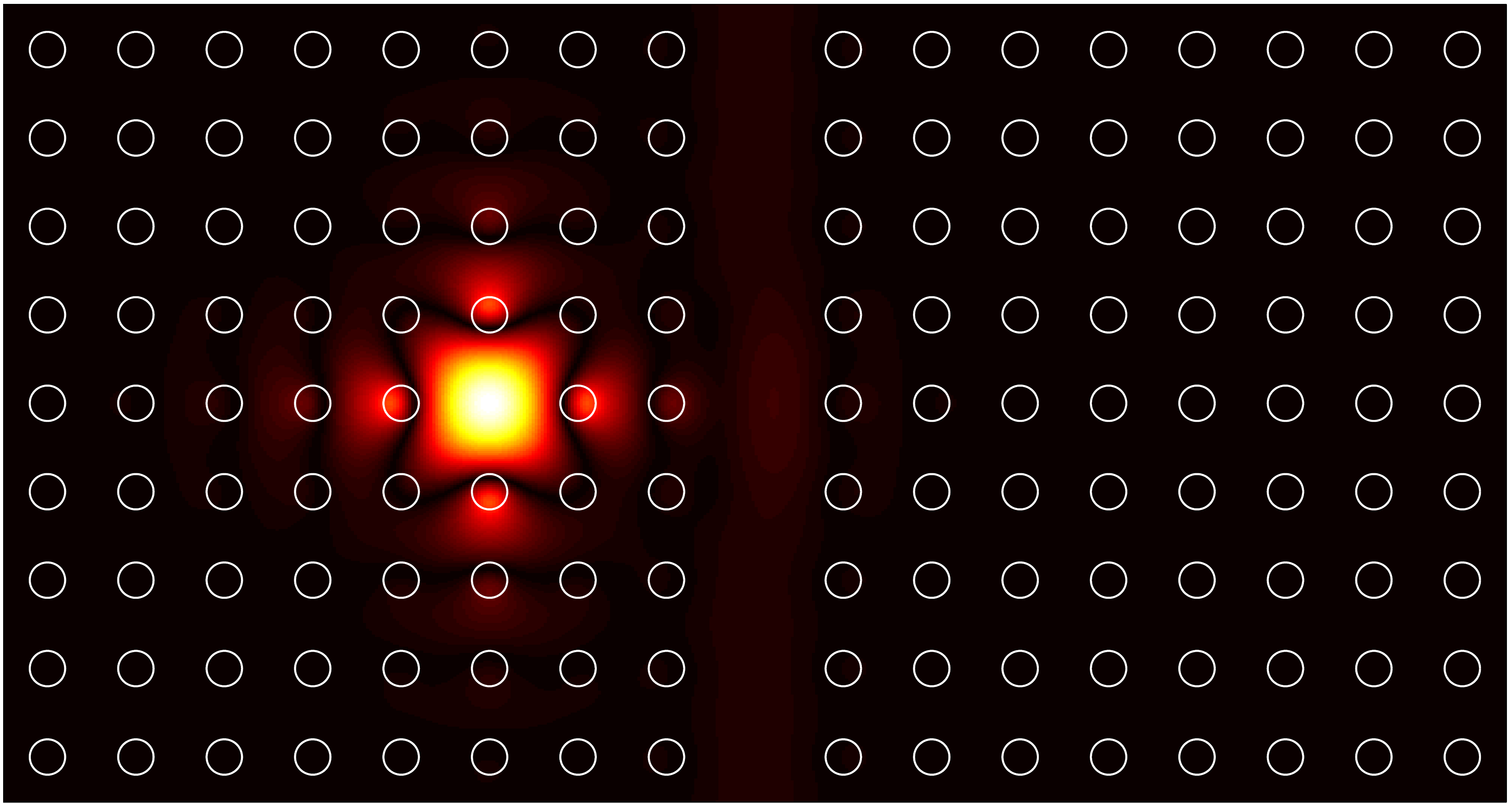}};

\draw (5.02-0.2-0.8,0.1+1.8) [color=red] arc (0:180:0.16cm); 
\draw [->,line width=0.1pt,color=red] (4.70-0.2-0.8,0.1+1.8) -- (4.70-0.2-0.8,-0.3+1.8) node(xline) at (4.8-0.2-0.8,-0.6+1.8) {\small{$R$}};

\draw [-,line width=0.5pt,color=red] (4.70-4.3-1.23,-3.5) -- (4.70-4.3-1.23,-3.80);
\node[anchor=west] at (4.7-4.65-1.0,-3.6) (text) {};
\node[anchor=south,color=red] at (4.25-1.2-7.7,-3.25) (description) {$\omega_{\mathrm{R}}$};
\draw[color=red,->] (description) to [out = 0, in = 120] (text);

\draw [-,line width=0.5pt,color=red] (4.70-4.45-1.32,0.3) -- (4.70-3.97-1.37,0.3);
\node[anchor=west] at (4.7-4.45-1.32,0.5) (text) {};
\node[anchor=south,color=red] at (4.25-8.82,-2.47) (description) {FWHM};
\draw[color=red,->] (description) to [out = 0, in = -90] (text);

\end{tikzpicture}

\caption{(Color online) Spectrum of power reflection $R$ of propagating Bloch mode in W1 defect waveguide in 2D rectangular PhC lattice. The inset shows the structure and the field distribution at the reflection maximum. The reflection is due to destructive interference between the propagating waveguide field and the cavity mode. The black crosses are obtained from scattering calculations, whereas the blue solid curve is obtained from Eq.~\eqref{Eqn:RQNM} and the complex frequency of the associated QNM. More details on the structure and modeling parameters are given in Section~\ref{Sec:QNMsPhCs} and Table~\ref{Tab:PhCParameters}.} \label{Fig:PhC_Reflection} 
\end{figure}

As a typical frequency domain calculation, Fig.~\ref{Fig:PhC_Reflection} shows the power reflection spectrum of the propagating Bloch mode of a W1 defect waveguide in a two-dimensional (2D) PhC due to a side-coupled defect cavity. These types of coupled PhC-waveguide-cavity structures are promising candidates for realization of optical switching~\cite{Nozaki2010,Rephaeli2012,Heuck2013a,Yu2013,Fan2002}. The inset shows the structure and the field distribution at the reflection maximum, and the excitation of a resonant cavity mode is clearly visible inside the cavity. The $Q$ factor is often deduced from this type of scattering calculation as~\cite{Joannopoulos2011Chap7}
\begin{align} \label{Eqn:QScatCalc}
Q = \dfrac{\omega_{\mathrm{R}}}{\mathrm{FWHM}},
\end{align}
where $\omega_{\mathrm{R}}$ and FWHM are the center frequency and the full width at half maximum of the Lorentzian spectrum, respectively, as indicated in Fig.~\ref{Fig:PhC_Reflection}. This follows from the assumption of a time-decaying resonant mode, whose electric field at a fixed position inside the cavity behaves as a damped harmonic oscillator in the time domain
\begin{align}\label{Eqn:EnergyDH}
\vekn{E}(t) = \vekn{E}_0\exp\left(- \ci \omegat t \right), \hspace{0.5cm} \omegat \equiv \omega_{\mathrm{R}} - \ci \gamma,  \hspace{0.25cm} \gamma > 0,
\end{align}
where $\omegat$ is a \textit{complex-valued} frequency.

An alternative description exists, in which the resonant modes are calculated explicitly as the time-harmonic solutions of the source-free Maxwell equations that satisfy an outgoing wave boundary condition (BC) (or radiation condition). This choice of BC renders the frequency domain wave equation problem non-Hermitian and leads to a discrete spectrum of complex mode frequencies $\omegat$. These solutions are the so-called quasi-normal modes (QNMs). Via the harmonic time dependence (see Eq.~\eqref{Eqn:EnergyDH}) this corresponds to \textit{temporally decaying} modes and thus to an explicit description of the leaky nature of the resonance, with the  $Q$ factor given as~\cite{Novotny2012Chap11}
\begin{align}
Q = \frac{\omega_{\mathrm{R}}}{2\gamma}.
\end{align}
In turn, the modes are \textit{spatially diverging}, describing ``the propagating front of the decaying state"~\cite{Tikhodeev2002}. The spatial divergence of the QNMs renders them non-trivial to normalize, but a rigorous normalization scheme was developed in~\cite{Leung1994,Lee1999} for simple systems. This result was used in~\cite{Kristensen2012} to introduce a generalized effective mode volume for leaky optical cavities, and in \cite{Sauvan2013} Sauvan \textit{et al.} elegantly employed the reciprocity theorem to derive an alternative formulation of the normalization and mode volume in dispersive media. These formulations have been shown to be equivalent~\cite{Ge2013}. Recently, a practical scheme for normalizing QNMs was proposed in~\cite{Bai2013}. QNMs have been used for modeling of several resonant nanophotonic structures: In~\cite{Settimi2003,Settimi2009}, QNMs were calculated and used for analysis of one-dimensional (1D) finite-sized PhCs, in~\cite{Sauvan2013} decay rates of dipole emitters in the vicinity of three-dimensional (3D) plasmonic resonantors were predicted accurately using QNMs, in~\cite{Bai2013} QNMs were applied to study scattering properties of propagating fields (plane waves, for example), and in~\cite{deLasson2013} QNMs were determined and discussed as localized surface plasmons of 3D plasmonic dimers. All these examples of QNMs from the literature pertain to finite-sized structures surrounded by homogeneous environments. In this article, we determine QNMs of cavities coupled to infinitely periodic PhC waveguides. For more details on QNMs in general physical systems see the review in~\cite{Ching1998}, and for a recent review on the use of QNMs in nanophotonics and plasmonics see~\cite{Kristensen2014}.

An intricate part of QNM calculations is to satisfy the outgoing wave BC. For spatial discretization techniques, like the finite-difference time-domain (FDTD) method~\cite{Taflove2005} and the finite element method (FEM)~\cite{Reddy2005}, this BC is notoriously difficult to implement and is typically approximated using so-called perfectly matched layers (PMLs)~\cite{Berenger1994}. In contrast, Green's function techniques can lead to solutions satisfying the outgoing wave BC analytically, and the QNMs can be determined as non-trivial solutions of ``excitation-free" volume~\cite{deLasson2013,Kristensen2012} or surface~\cite{Makitalo2014} integral equations. Green's function techniques are well-suited for treating the outgoing wave BC, but in turn have difficulties in handling large complex-shaped geometries like the PhC membrane. As an alternative, modal expansion and scattering matrix techniques~\cite{Lavrinenko2014Chap6} have been effectively used for treating these kinds of structures. The total scattering matrix $\vekn{S}$ relates the outgoing to the incoming fields, and the QNMs can be defined as the non-trivial output for vanishing input
\begin{align} \label{Eqn:QNMConditionLit0}
|\mathrm{out} \rangle = \vekn{S} |\mathrm{in} \rangle, \,\,\,\,\, |\mathrm{in} \rangle \rightarrow 0,
\end{align}
with solutions of Eq.~\eqref{Eqn:QNMConditionLit0} existing at complex frequencies where $\vekn{S}$ has a pole~\cite{Petit1980Chap5}. This was used for determining QNMs in PhC slabs in~\cite{Tikhodeev2002}, and pole expansions of the scattering matrix have been proposed for determining QNMs in both PhC~\cite{Akimov2011} and metallic~\cite{Bykov2013} slabs. Similarly, a linearization of the inverse scattering matrix in connection with the Fourier Modal Method (FMM) has been developed for determining QNMs of periodic metallic systems~\cite{Weiss2011a}. However, for complex geometries the scattering matrix is comparatively large, and inverting this large and singular matrix is computationally demanding. 

In this article, we propose a simple technique for determining QNMs using scattering matrices. The method is based on the definition of an internal cavity section and an iteration of the complex frequency towards a unity eigenvalue of the associated cavity roundtrip matrix. This is equivalent to the well-known laser oscillation condition, employed when analyzing lasers, and similar procedures have been used to analyze and design micropillar cavity structures~\cite{Lecamp2005,Gregersen2010}. Practically, the method is advantageous since it avoids the numerical calculation of an inverse scattering matrix. We develop the method using a Bloch mode expansion technique for periodic structures, but the procedure can be used with any routine based on scattering matrices. The method can be used for determining QNMs of finite-sized structures, but by use of Bloch mode expansions it can in particular also be used for determining QNMs in infinitely periodic structures, which we demonstrate for two types of PhC cavities coupled to waveguides. 

The article is organized as follows: Section~\ref{Sec:QNMScatMatrices} introduces the modal expansion and scattering matrix techniques for determining QNMs. The generic form of the structure as treated with these methods is presented, and the Bloch mode expansions are introduced. The mathematical definition of the QNMs is given, and finally the method for determining QNMs based on the inverse scattering matrix is reviewed, and the new method based on the cavity roundtrip matrix is presented. In Section~\ref{Sec:QNMsPhCs}, example calculations of QNMs using the cavity roundtrip matrix method are provided. Specifically, we determine QNMs of two types of 2D PhC cavity structures and discuss their $Q$ factors and spatial distributions. Section~\ref{Sec:Conclusion} concludes the work. Appendices~\ref{App:ClassBlochModes} and \ref{App:EquivalenceMethods} provide additional details on the determination of Bloch modes and on the equivalence between the scattering matrix and roundtrip matrix methods, respectively.

\section{Bloch mode expansions, scattering matrices, and quasi-normal modes} \label{Sec:QNMScatMatrices}
In this section, we present the computational framework for calculating QNMs using a cavity roundtrip matrix. In Section~\ref{Sec:FMMIntro}, we introduce the generic form of the structures and present the Bloch mode expansions of the electromagnetic fields. Having introduced the general framework, we use Section~\ref{Sec:QNMTheory} to state the equations satisfied by QNMs, in particular for the infinitely periodic structures to be analyzed in Section~\ref{Sec:QNMsPhCs}. Finally, Section~\ref{Sec:QNMMethods} reviews the methods for calculating the QNMs based on the total scattering matrix and outlines the details of the new method for determining the QNMs using an internal cavity and its roundtrip matrix.

\subsection{Bloch mode expansions and scattering matrices}\label{Sec:FMMIntro}
\begin{figure}[htb!]
\centering
\vspace{-1cm}

\begin{tikzpicture} [line cap=round,line join=round,>=triangle 45,x=1.0cm,y=1.0cm, scale=0.47]

\newcommand{\dx}{-4}
\newcommand{\dxq}{4}
\newcommand{\dyq}{-0.75}

\draw [->] (\dx+2,-2)  -- (\dx+3,-2) node(xline)[right]{$\vekr_\perp$};
\draw [->] (\dx+2,-2)  -- (\dx+2,-1) node(yline)[right]{$z$}; 

\fill[pattern=north west lines, pattern color=black!50!white] (\dx-5,1.5) rectangle (\dx+1.5,3);
\fill[pattern=north east lines, pattern color=blue!50!white] (\dx-5,3) rectangle (\dx+1.5,4);
\fill[pattern=north west lines, pattern color=black!50!white] (\dx-5,4) rectangle (\dx+1.5,5.5);
\fill[pattern=north east lines, pattern color=blue!50!white] (\dx-5,5.5) rectangle (\dx+1.5,6.5);

\draw[-,color=black,line width=1pt] (\dx-5,0.5) -- (\dx+1.5,0.5);
\draw[-,color=black,line width=1pt] (\dx-5,1.5) -- (\dx+1.5,1.5);
\draw[-,color=black,line width=1pt] (\dx-5,3) -- (\dx+1.5,3);
\draw[-,color=black,line width=1pt] (\dx-5,4) -- (\dx+1.5,4);
\draw[-,color=black,line width=1pt] (\dx-5,5.5) -- (\dx+1.5,5.5);
\draw[-,color=black,line width=1pt] (\dx-5,6.5) -- (\dx+1.5,6.5);
\draw[-,color=black,line width=1pt] (\dx-5,7.5) -- (\dx+1.5,7.5);

\draw (\dx-3,-0.25) node[circle]{Layer 1};
\draw (\dx+0,1.25) node[circle]{$\vdots$};
\draw (\dx-3,2.25) node[circle]{Layer $q-1$};
\draw (\dx-3,3.55) node[circle]{Layer $q$};
\draw (\dx-3,4.75) node[circle]{Layer $q+1$};
\draw (\dx-3,6) node[circle]{Layer $q+2$};
\draw (\dx+0,7.25) node[circle]{$\vdots$};
\draw (\dx-3,8.25) node[circle]{Layer $Q$};

\draw[-,color=black,dashed,line width=1pt] (\dxq-5.15,5.5+\dyq) -- (\dx+1.6,6.5);
\draw[-,color=black,dashed,line width=1pt] (\dxq-5.15,4+\dyq) -- (\dx+1.6,1.5);

\draw (\dx-1.5,9.5) node[circle]{\textbf{Eigenmodes:}};
\draw (\dxq-1.5,9.5) node[circle]{\textbf{Bloch Modes:}};

\draw[-,color=black,line width=1pt] (\dxq-5,1.5+\dyq) -- (\dxq+3.5,1.5+\dyq);
\draw[-,color=black,line width=1pt] (\dxq-5,3+\dyq) -- (\dxq+3.5,3+\dyq);
\draw[-,color=black,line width=1pt] (\dxq-5,4+\dyq) -- (\dxq+3.5,4+\dyq);
\draw[-,color=black,line width=1pt] (\dxq-5,5.5+\dyq) -- (\dxq+3.5,5.5+\dyq);
\draw[-,color=black,line width=1pt] (\dxq-5,6.5+\dyq) -- (\dxq+3.5,6.5+\dyq);

\draw (\dxq-3,0.75+\dyq) node[circle]{Section 1};
\draw (\dxq-3,2.25+\dyq) node[circle]{Section 2};
\draw (\dxq+0,3.75+\dyq) node[circle]{$\vdots$};
\draw (\dxq-3,4.75+\dyq) node[circle]{Section $w$};
\draw (\dxq+0,6.25+\dyq) node[circle]{$\vdots$};
\draw (\dxq-3,7.25+\dyq) node[circle]{Section $W$};

\draw[->,color=blue,line width=1pt] (\dxq+1,-0.5+\dyq) -- (\dxq+1,1.3+\dyq);
\draw[->,color=blue,line width=1pt] (\dxq+3,8.5+\dyq) -- (\dxq+3,6.7+\dyq);
\draw[->,color=red,dashed,line width=1pt] (\dxq+1,6.7+\dyq) -- (\dxq+1,8.5+\dyq);
\draw[->,color=red,dashed,line width=1pt] (\dxq+3,1.3+\dyq) -- (\dxq+3,-0.5+\dyq);
\draw (\dxq+0,-2+\dyq) node[circle]{$\textcolor{red}{\vekn{c}_{\mathrm{out}}} = \vekn{S} \textcolor{blue}{\vekn{c}_{\mathrm{in}}}$};

\draw [decorate,decoration={brace,amplitude=10pt,mirror},xshift=0pt,yshift=0pt]
(\dx-5.2,4) -- (\dx-5.2,1.5) node [black,midway,xshift=-0.6cm,yshift=0cm] {\rotatebox{90}{Supercell}};

\end{tikzpicture}
\vspace{-1cm}
\caption{(Color online) Generic structure as analyzed using modal expansions. \textbf{Left panel:} Layered structure consisting of $Q$ $z$ invariant layers. Hatching patterns indicate lateral permittivity profiles, and layers $q-1$ and $q+1$ ($q$ and $q+2$) are thus identical.  \textbf{Right panel:} Layered structure consisting of $W$ periodic sections that each comprises a number of layers. Layers $q-1$ through $q+2$ (left panel) constitute the $w$th periodic section (right panel) with two repetitions of the supercell. The amplitudes of the incoming Bloch modes in sections 1 and $W$ (solid arrows) are related to the amplitudes of the outgoing Bloch modes (dashed arrows) via the total scattering matrix $\vekn{S}$.} \label{Fig:GeometryFMM} 
\end{figure}
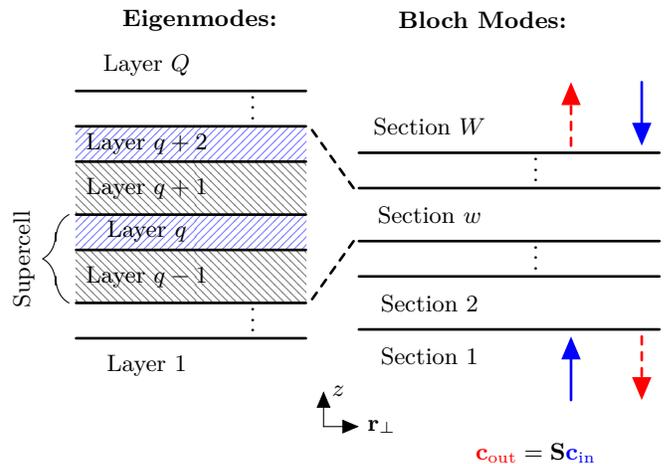

In the modal expansion techniques, the structure under consideration is divided into layers, which are uniform along a chosen propagation direction, taken here as the $z$ axis. In each layer, so-called lateral eigenmodes are determined as solutions to the $z$ invariant wave equation~\cite{Lavrinenko2014Chap6}. A generic sketch of the structure consisting of $Q$ layers is shown in the left panel of Fig.~\ref{Fig:GeometryFMM} where hatching patterns indicate lateral permittivity profiles; layers $q-1$ and $q+1$ ($q$ and $q+2$) are identical. The mode profiles of the lateral eigenmodes can in certain symmetric cases be expressed analytically~\cite{Snyder1983}, and otherwise finite-difference or Fourier series~\cite{Noponen1994,Moharam1995} techniques are used. The lateral eigenmodes are used as a basis for expansion of the electric and magnetic fields, and by satisfying the electromagnetic BCs at interfaces, scattering matrices of the entire multi-layer structure are determined using an iterative procedure~\cite{Li1996}.

Structures exhibiting a periodicity in the propagation direction $z$ may be analyzed more intuitively and efficiently by imposing Bloch's theorem~\cite{Joannopoulos2011Chap3}. Instead of constructing the scattering matrix for the entire structure, the scattering matrix of a single supercell is considered, and the characteristic modes of each supercell are the Bloch modes. In Fig.~\ref{Fig:GeometryFMM}, layers $q-1$ through $q+2$ (left panel) constitute the $w$th periodic section (right panel), with layers layers $q-1$ and $q$ being the supercell. The grouping of layers gives rise to a conceptually new structure consisting of $W$ periodic sections, as shown in the right panel of Fig.~\ref{Fig:GeometryFMM}. In the sectioned structure, the fields in each section $w$, $\vekE^w(\vekr)$ and $\vekH^w(\vekr)$, are expanded on the Bloch modes, $\vekn{e}_j^w(\vekr_{\perp},z)$ and $\vekn{h}_j^w(\vekr_{\perp},z)$,
\begin{subequations} \label{Eqn:FieldExpansions}
\begin{align} \label{Eqn:FieldExpansionE}
\vekE^w(\vekr) &= \sum_j c_j^w \vekn{e}_j^w(\vekr_{\perp},z), \\
\vekH^w(\vekr) &= \sum_j c_j^w \vekn{h}_j^w(\vekr_{\perp},z), \label{Eqn:FieldExpansionH}
\end{align}
\end{subequations}
which, owing to Bloch's theorem, are quasi-periodic functions of the $z$ coordinate,
\begin{subequations} \label{Eqn:BlochModes}
\begin{align} \label{Eqn:BlochModeE}
\vekn{e}_j^w(\vekr_{\perp},z+a^w) &= \rho_j^w \vekn{e}_j^w(\vekr_{\perp},z), \\
\label{Eqn:BlochModeH}
\vekn{h}_j^w(\vekr_{\perp},z+a^w) &= \rho_j^w \vekn{h}_j^w(\vekr_{\perp},z), \\
\rho_j^w &\equiv \exp(\ci k_{j}^{w} a^w), \label{Eqn:BlochModeRho}
\end{align}
\end{subequations}
where $k_{j}^w$ and $a^w$ are the wavenumber of the $j$th Bloch mode and the length of the supercell, respectively. We refer to $\rho_j^w$ as the Bloch mode eigenvalue, and this is the central quantity that must be determined; when it is known, Eqs.~\eqref{Eqn:FieldExpansions} and \eqref{Eqn:BlochModes} provide an analytic description of the propagation of each Bloch mode in the $z$ direction without the need to impose artificial BCs like PMLs. This represents a major advantage and is an important element in satisfying the outgoing wave BC of the QNMs with the present approach. In Appendix~\ref{App:ClassBlochModes}, we detail the technique for calculating the Bloch modes and discuss their classification as either decaying, propagating or growing in the upward ($+z$) or downward ($-z$) directions. Bloch mode reflection and transmission matrices at section interfaces are derived by requiring continuity of the tangential components of the fields, and subsequently Bloch mode scattering matrices are derived iteratively. This allows for the construction of fields expanded on Bloch modes in all sections for an arbitrary excitation. Importantly, we may write Eq.~\eqref{Eqn:QNMConditionLit0} in terms of the Bloch mode amplitudes introduced in Eqs.~\eqref{Eqn:FieldExpansions} to relate the amplitudes of the incoming $\vekn{c}_{\mathrm{in}}$ and outgoing $\vekn{c}_{\mathrm{out}}$ Bloch modes in the outermost regions (solid and dashed arrows, respectively, in the right panel of Fig.~\ref{Fig:GeometryFMM}) through the total scattering matrix $\vekn{S}$
\begin{align} \label{Eqn:ScatMatrix}
\vekn{c}_{\mathrm{out}} = \vekn{S} \vekn{c}_{\mathrm{in}}.
\end{align}
This relation forms the basis of the scattering matrix method for determining QNMs. We stress that for the determination of QNMs as presented in Section~\ref{Sec:QNMMethods}, the use of scattering matrices is crucial, while the use of Bloch mode expansions is simply our specific choice. Alternatives include Green's function based expansions~\cite{Botten2000} in connection with scattering matrices and Bloch modes~\cite{Botten2001}.

\subsection{Definition of quasi-normal modes} \label{Sec:QNMTheory}
We define QNMs as time-harmonic solutions $\vekn{E}(\vekr; t) = \vekn{E}(\vekr; \omegat) \exp \left(- \ci \omegat t\right)$ with the frequency component $\vekn{E}(\vekr; \omegat)$ satisfying the wave equation
\begin{align}\label{Eqn:EWaveEqn}
\nabla \times \nabla \times \vekn{E}(\vekr; \omegat) - k_0^2 \epsilon(\vekr; \omegat) \vekn{E}(\vekr; \omegat) = 0,
\end{align}
and an outgoing wave BC (or radiation condition)~\cite{Kristensen2014}, to be specified below. In Eq.~\eqref{Eqn:EWaveEqn}, $k_0 = \omegat/\mathrm{c}$ and $\epsilon(\vekr; \omegat)$ are the free-space wave number and the relative permittivity, respectively. For structures of finite extent and embedded in a homogeneous environment of refractive index $n_{\mathrm{B}}$, the outgoing wave BC is the so-called Silver-M\"{u}ller radiation condition~\cite{Martin2006}
\begin{align} \label{Eqn:SilverMuller}
\hat{\vekr} \times \nabla \times \vekn{E}(\vekr; \omegat) + \ci n_{\mathrm{B}} k_0 \vekn{E}(\vekr; \omegat) \rightarrow 0 \,\,\, \mathrm{for} \,\,\, n_{\mathrm{B}} k_0 |\vekr| \rightarrow \infty.
\end{align}
This choice of BC, as mentioned in Section~\ref{Sec:Introduction}, renders the differential equation problem non-Hermitian and thus in general leads to solutions with complex frequencies. In the context of the modal expansions of Eqs.~\eqref{Eqn:FieldExpansions}, we formulate the outgoing wave BC of the QNMs as a finite output at vanishing input
\begin{align} \label{Eqn:QNMBC}
\vekn{c}_{\mathrm{out}} \neq \vekn{0} \,\,\, \mathrm{for} \,\,\, \vekn{c}_{\mathrm{in}} = \vekn{0}.
\end{align}
If the outermost sections 1 and $W$ (see the right panel in Fig.~\ref{Fig:GeometryFMM}) are homogeneous, the condition in Eq.~\eqref{Eqn:QNMBC} is equivalent to the radiation condition in Eq.~\eqref{Eqn:SilverMuller}. If these sections are not homogeneous, the Silver-M\"{u}ller radiation condition in Eq.~\eqref{Eqn:SilverMuller} cannot be applied, while the condition in Eq.~\eqref{Eqn:QNMBC} remains a viable condition, and we use this to determine QNMs in PhC cavities coupled to infinitely extended waveguides in Section~\ref{Sec:QNMsPhCs}. 

\subsection{Calculating quasi-normal modes using scattering matrices}\label{Sec:QNMMethods}
In this section, we first review the method based on the total scattering matrix in Section~\ref{Sec:ScatteringMethod}. Thereafter we present the new method based on an internal cavity and its roundtrip matrix in Section~\ref{Sec:RoundtripMethod}. We term these the total scattering matrix method and the cavity roundtrip matrix method, respectively. 

\subsubsection{Total scattering matrix method}\label{Sec:ScatteringMethod}
The total scattering matrix method relies on solving Eq.~\eqref{Eqn:ScatMatrix} explicitly with the condition in Eq.~\eqref{Eqn:QNMBC}, which becomes
\begin{align} \label{Eqn:QNMConditionLit1}
\vekn{S}^{-1}(\omegat) \vekn{c}_{\mathrm{out}} = \vekn{0}.
\end{align}
QNMs are the non-trivial solutions of this equation, leading to the following condition for the QNM frequency
\begin{align} \label{Eqn:QNMConditionLit2}
\mathrm{det}\left(\vekn{S}^{-1}(\omegat) \right) = 0.
\end{align}
This equation, in general, is satisfied at complex-valued $\omegat$, and the associated non-trivial $\vekn{c}_{\mathrm{out}}$ yields the expansion coefficients of the QNM on the Bloch modes in the outermost sections. This, in principle, provides a straightforward method for determining QNMs. In our experience, however, the solution of Eqs.~\eqref{Eqn:QNMConditionLit1} and \eqref{Eqn:QNMConditionLit2} is complicated; for complex structures that require the inclusion of a large number of Bloch modes, the scattering matrix is comparatively large and the construction of its inverse matrix becomes numerically inaccurate and unstable. This can be understood as follows: Near the QNM complex frequency, some elements in the scattering matrix $\vekn{S}$ become very large to ensure a finite output at vanishing input. At the same time other elements in the scattering matrix remain small, and it therefore contains elements spanning many orders of magnitude, rendering the numerical construction of the inverse scattering matrix infeasible. To address this issue it has been proposed to employ the singular nature of the scattering matrix close to a QNM frequency by expanding the singular part of the matrix in simple poles around the complex mode frequency, and we refer the reader to~\cite{Tikhodeev2002,Weiss2011a,Akimov2011,Bykov2013} for details on this procedure. In certain simple cases, the contributions to the QNM are dominated by a single Bloch mode, and it is possible to locate the QNM spectrally by looking for a pole of the scattering coefficient of this single Bloch mode similarly to what has been done for grating structures~\cite{Petit1980Chap5}.

\subsubsection{Cavity roundtrip matrix method}\label{Sec:RoundtripMethod}
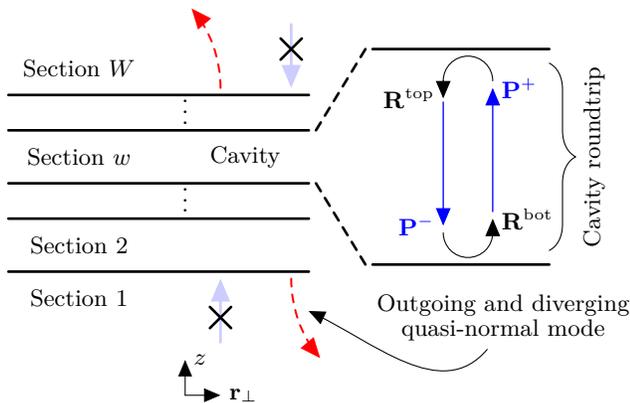
\begin{figure}[htb!]
\centering

\begin{tikzpicture} [line cap=round,line join=round,>=triangle 45,x=1.0cm,y=1.0cm, scale=0.47]

\newcommand{\dx}{0}
\newcommand{\dxq}{0}

\draw [->] (\dx-0,-2)  -- (\dx+1,-2) node(xline)[right]{$\vekr_\perp$};
\draw [->] (\dx-0,-2)  -- (\dx-0,-1) node(yline)[right]{$z$}; 

\draw[-,color=black,line width=1pt] (\dx-5,1.5) -- (\dx+3.5,1.5);
\draw[-,color=black,line width=1pt] (\dx-5,3) -- (\dx+3.5,3);
\draw[-,color=black,line width=1pt] (\dx-5,4) -- (\dx+3.5,4);
\draw[-,color=black,line width=1pt] (\dx-5,5.5) -- (\dx+3.5,5.5);
\draw[-,color=black,line width=1pt] (\dx-5,6.5) -- (\dx+3.5,6.5);

\draw (\dx-3,0.75) node[circle]{Section 1};
\draw (\dx-3,2.25) node[circle]{Section 2};
\draw (\dx+0,3.75) node[circle]{$\vdots$};
\draw (\dx-3,4.75) node[circle]{Section $w$};
\draw (\dx+1.7,4.75) node[circle]{Cavity};
\draw (\dx+0,6.25) node[circle]{$\vdots$};
\draw (\dx-3,7.25) node[circle]{Section $W$};

\draw[->,color=blue!20!white,line width=1pt] (\dx+1,-0.5) -- (\dx+1,1.3);
\draw[->,color=blue!20!white,line width=1pt] (\dx+3,8.5) -- (\dx+3,6.7);

\draw [->,color=red,dashed,line width=0.75pt] (\dx+1,6.7) arc (0:40:3.5cm); 
\draw [->,color=red,dashed,line width=0.75pt] (\dx+3,1.3) arc (180:220:3.5cm);

\node[anchor=west] at (\dx+3,0.5) (text) {};
\node[anchor=south] at (\dx+9,0) (description) {Outgoing and diverging};
\node[anchor=south] at (\dx+9,-0.7) (description) {quasi-normal mode};
\draw[color=black,->] (description) to [out = -130, in = -25] (text);

\draw[-,color=black,line width=1pt] (\dx+1.3,0.5) -- (\dx+0.7,-0.1);
\draw[-,color=black,line width=1pt] (\dx+0.7,0.5) -- (\dx+1.3,-0.1);
\draw[-,color=black,line width=1pt] (\dx+1.3+2,0.5+7.6) -- (\dx+0.7+2,-0.1+7.6);
\draw[-,color=black,line width=1pt] (\dx+0.7+2,0.5+7.6) -- (\dx+1.3+2,-0.1+7.6);

\draw[-,color=black,dashed,line width=1pt,dash pattern=on 3.5pt off 3pt] (\dx+3.7,5.5) -- (\dx+5.1,7.8);
\draw[-,color=black,dashed,line width=1pt,dash pattern=on 3.5pt off 3pt] (\dx+3.7,4) -- (\dx+5.1,1.7);

\draw[-,color=black,line width=1pt] (\dx+5.3,7.8) -- (\dx+10.3,7.8);
\draw[-,color=black,line width=1pt] (\dx+5.3,1.7) -- (\dx+10.3,1.7);

\draw (\dx+8.7,6.9) arc (0:180:0.7cm); 
\draw [->,line width=0.5pt] (\dx+7.3,6.9) -- (\dx+7.3,6.4) node(xline) [left] {$\vekn{R}^{\mathrm{top}}$};

\draw (\dx+7.3,2.6) arc (-180:0:0.7cm); 
\draw [->,line width=0.5pt] (\dx+8.7,2.6) -- (\dx+8.7,3.0) node(xline)[right]{$\vekn{R}^{\mathrm{bot}}$};

\draw [->,line width=0.5pt,color=blue] (\dx+8.7,3.2) -- (\dx+8.7,6.7) node(xline)[right]{$\vekn{P}^{+}$};

\draw [->,line width=0.5pt,color=blue] (\dx+7.3,6.3) -- (\dx+7.3,2.8) node(xline)[left]{$\vekn{P}^{-}$};

\draw [decorate,decoration={brace,amplitude=10pt},xshift=0pt,yshift=0pt]
(\dx+10.3,7.4) -- (\dx+10.3,2.1) node [black,midway,xshift=+0.6cm] {\rotatebox{90}{Cavity roundtrip}};

\end{tikzpicture}

\caption{(Color online) Outline of the cavity roundtrip matrix method for determining QNMs. \textbf{Left panel:} In the sectioned structure introduced in Fig.~\ref{Fig:GeometryFMM} an internal section $w$ is chosen as the cavity. In the outermost regions, the amplitudes of the incoming (solid arrows) and outgoing (dashed arrows) Bloch modes vanish and have finite values, respectively. The outgoing modes grow and diverge as they propagate away from the structure. \textbf{Right panel:} Zoom in on cavity section with indication of the elements of the roundtrip matrix comprising the top and bottom scattering reflection matrices, $\vekn{R}^{\mathrm{top}}$ and $\vekn{R}^{\mathrm{bot}}$, and up- and downward propagation matrices, $\vekn{P}^{+}$ and $\vekn{P}^{-}$.} \label{Fig:GeometryCavityRoundtrip} 
\end{figure}

To introduce the cavity roundtrip matrix method, we consider the sectioned structure, as introduced in the right panel of Fig.~\ref{Fig:GeometryFMM}, and select an internal section $w$, $2 \leq w \leq W-1$, that we refer to as the cavity section, see the left panel of Fig.~\ref{Fig:GeometryCavityRoundtrip}. An example of a cavity is section 2 in Fig.~\ref{Fig:QNM_SideCouple}(a), and in Section~\ref{Sec:PhCInlineQNM} we discuss the choice of cavity section further.

In the cavity section, we search for the QNMs as superpositions of Bloch modes $\vekn{c}_{\mathrm{c}}$ that replicate themselves upon a roundtrip
\begin{align} \label{Eqn:RoundtripMatrixEqn}
\vekn{M} \vekn{c}_{\mathrm{c}} = \alpha_{\mathrm{c}} \vekn{c}_{\mathrm{c}}, \,\,\,\,\, \alpha_{\mathrm{c}} = 1, 
\end{align}
where $\alpha_{\mathrm{c}}$ is an eigenvalue of the cavity roundtrip matrix $\vekn{M}$~\cite{Gregersen2010}. $\vekn{M}$ is obtained by multiplying the reflection and propagation matrices in the order indicated by the arrows in the right panel of Fig.~\ref{Fig:GeometryCavityRoundtrip}
\begin{align} \label{Eqn:RoundtripMatrix}
\vekn{M}(\omegat) \equiv \vekn{R}^{\mathrm{bot}} \vekn{P}^{-} \vekn{R}^{\mathrm{top}} \vekn{P}^{+},
\end{align}
where $\vekn{R}^{\mathrm{top}}$ ($\vekn{R}^{\mathrm{bot}}$) is the scattering reflection matrix for the top (bottom) part of the structure, while $\vekn{P}^{+}$ ($\vekn{P}^{-}$) is the diagonal matrix accounting for the propagation of the Bloch modes from the bottom (top) to the top (bottom) of the cavity section. In passive structures, the roundtrip eigenvalues $\alpha_{\mathrm{c}}$ always have magnitudes smaller than unity at real frequencies; part of the light leaks out of the cavity with reflectivities being smaller than unity. It is illustrative to compare to the case of lasers, where this loss is compensated by adding gain material, which gives elements with magnitudes larger than unity in the propagation matrices, and lasing starts when $\alpha_{\mathrm{c}} = 1$. In both passive and active structures, Eq.~\eqref{Eqn:RoundtripMatrixEqn} has solutions at complex frequencies, due to the radiation condition, and we solve it by iterating the frequency in the complex plane to find a cavity roundtrip eigenvalue equal to unity; the associated eigenvector $\vekn{c}_{\mathrm{c}}$ gives the QNM field distribution inside the cavity. Importantly, as indicated in the left panel of Fig.~\ref{Fig:GeometryCavityRoundtrip}, there are only outgoing waves in the outermost sections. This procedure therefore relies on construction of the cavity roundtrip matrix and determination of its eigenvalues, which is computationally more stable than inversion of the total scattering matrix. In Appendix~\ref{App:EquivalenceMethods}, we demonstrate that the two methods lead to the same equation for the complex QNM frequencies in a case where the scattering and roundtrip matrices can be expressed in closed form.

\section{Quasi-normal modes in photonic crystal cavities} \label{Sec:QNMsPhCs} 
To demonstrate the roundtrip matrix method proposed in Section~\ref{Sec:RoundtripMethod} we apply it to determine QNMs in two types of PhC cavities. For simplicity, we study 2D structures with uniformity along the $y$ direction in which case we may describe the electromagnetic fields completely by either the $y$ component of the electric $E_{y}$ (TE) or the magnetic field $H_{y}$ (TM). The lateral eigenmodes are determined using the FMM~\cite{Noponen1994,Moharam1995}, and Bloch modes in periodic sections are determined as detailed in Appendix~\ref{App:ClassBlochModes}. Note our convention of TE and TM, which is consistent with the FMM literature, but opposite of the convention used in the literature on PhCs~\cite{Joannopoulos2011Chap5}. 

\begin{table}
\centering
\begin{tabular}{c|c|c|c|c}
$r/a$ & $\epsilon_{\mathrm{Rods}}$ & $\epsilon_{\mathrm{Back}}$ & $N_{\mathrm{Fourier}}$ & $N_{\mathrm{Staircase}}$ \\
 \hline 
 0.2 & 8.9 & 1 & 101 & 128
\end{tabular}
\caption{Parameters of rectangular PhC lattice shown in inset in Fig.~\ref{Fig:PhC_Reflection} and in Figs.~\ref{Fig:QNM_SideCouple} and \ref{Fig:QNM_InLineCouple}; rod radius, $r$, and rod and background permittivity, $\epsilon_{\mathrm{Rods}}$ and $\epsilon_{\mathrm{Back}}$, respectively, as well as modeling parameters; number of Fourier series terms per layer, $N_{\mathrm{Fourier}}$, and number of steps in staircase approximation of rods, $N_{\mathrm{Staircase}}$.} \label{Tab:PhCParameters}
\end{table}

We consider a 2D rectangular PhC lattice (lattice constant $a$, see Fig.~\ref{Fig:QNM_SideCouple}) consisting of high-index rods (radius $r$, permittivity $\epsilon_{\mathrm{Rods}}$) in air (permittivity $\epsilon_{\mathrm{Back}}$), with parameters as specified in Table~\ref{Tab:PhCParameters}. This structure is known to possess a TE bandgap~\cite{Joannopoulos2011Chap5}, and a W1 waveguide is formed by removing a row of rods; this waveguide supports a propagating Bloch mode for guiding light through the structure. In sections~\ref{Sec:PhCSideQNM} and \ref{Sec:PhCInlineQNM}, we further introduce PhC cavities that are coupled to the W1 waveguide. The particular structures considered in Section~\ref{Sec:PhCSideQNM} are similar to those in~\cite{Fan2002}.

In each layer (left panel in Fig.~\ref{Fig:GeometryFMM}), we include a total of $N_{\mathrm{Fourier}}$ terms in the Fourier series for determining the lateral eigenmodes. By using the Bloch mode condition in the $z$ direction, see Eqs.~\eqref{Eqn:BlochModes}, we model a rectangular PhC lattice with a W1 waveguide extending infinitely along the $z$ direction without the need to employ absorbing BCs; along the $x$ direction we use periodic BCs. To meet the requirements of $z$ invariant layers (left panel in Fig.~\ref{Fig:GeometryFMM}), we approximate the rods of the PhC with a staircase consisting of $N_{\mathrm{Staircase}}$ layers per rod; we use values of $N_{\mathrm{Fourier}}$ and $N_{\mathrm{Staircase}}$ as given in Table~\ref{Tab:PhCParameters} and have verified that the results given in the following sections are not affected when increasing the values of these parameters.

\subsection{Cavity side-coupled to W1 waveguide} \label{Sec:PhCSideQNM}
\begin{figure}[htb!]
\centering 

\begin{tikzpicture} [line cap=round,line join=round,>=triangle 45,x=1.0cm,y=1.0cm, scale=0.47]
%


\node (img) at (0,0){\includegraphics[scale=0.19]{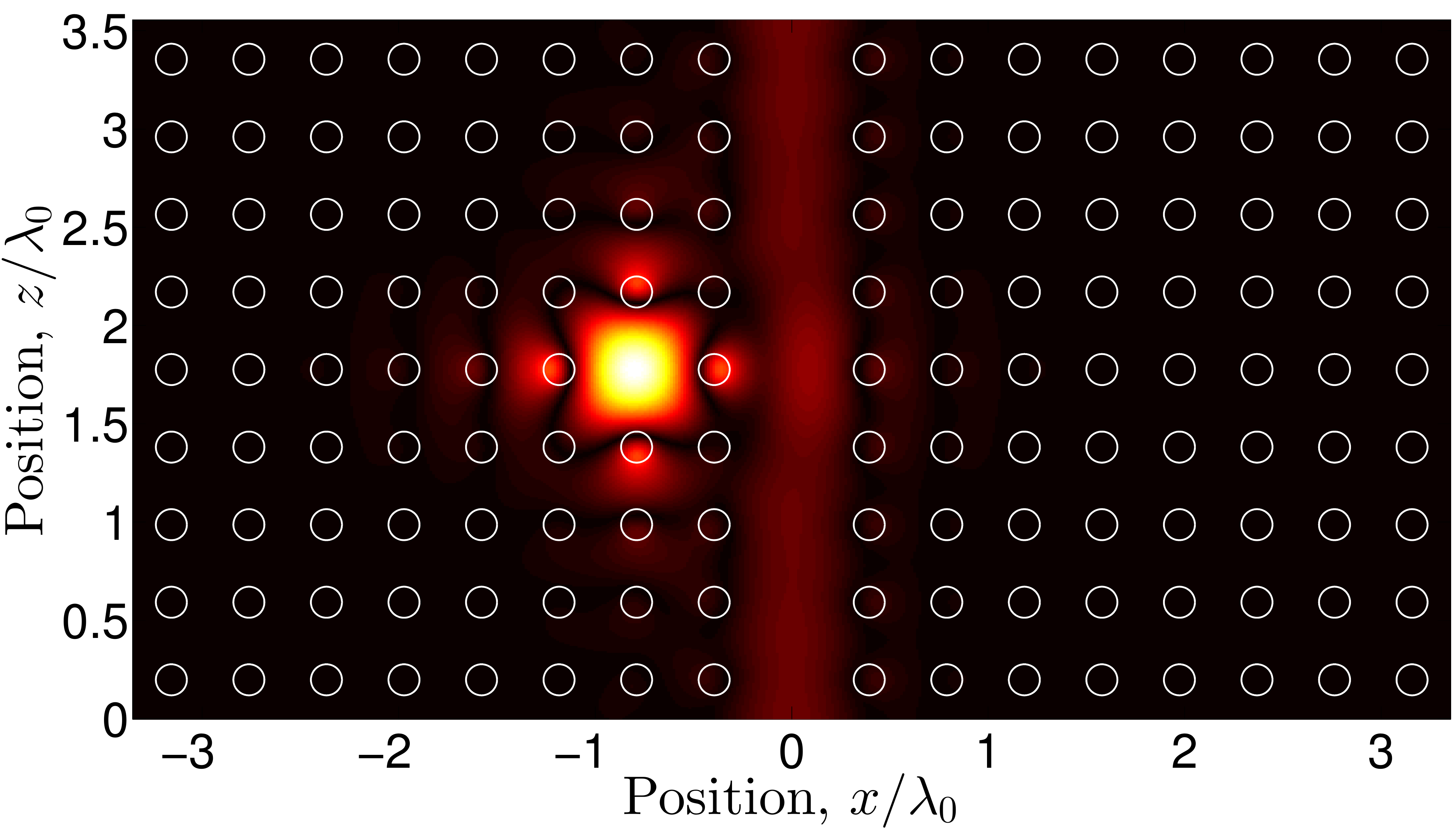}};
\node (img) at (0,-11){\includegraphics[scale=0.19]{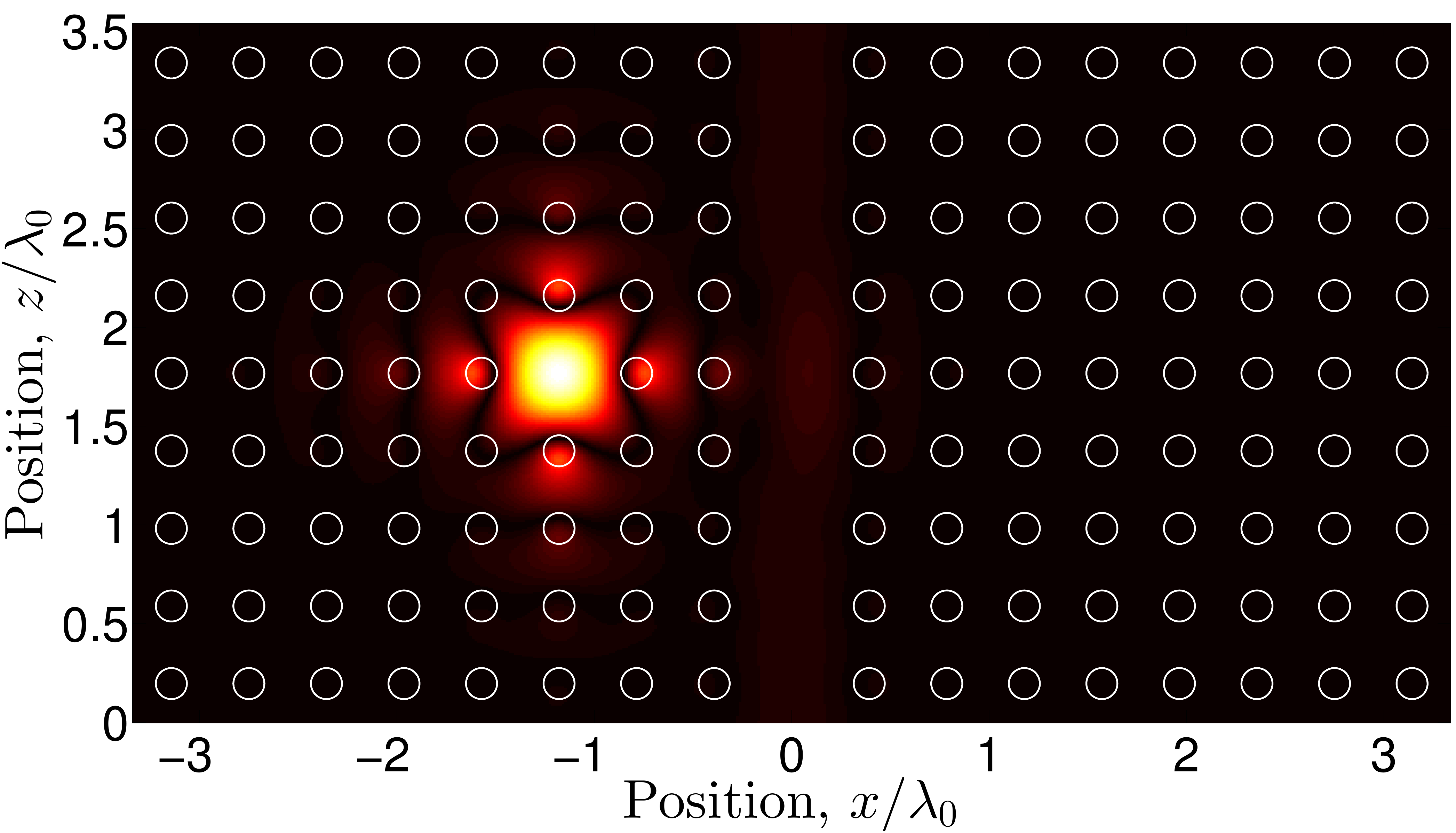}};

\draw [color=black] (-6.2,5.4) node[circle]{\large (a)};
\draw [color=black] (-6.2,-5.6) node[circle]{\large (b)};

\draw[-,white=white,line width=1pt,dashed] (-7.18+0.15,0.1) -- (8.75-0.2,0.1);
\draw[-,white=white,line width=1pt,dashed] (-7.18+0.15,1.0) -- (8.75-0.2,1.0);

\draw [color=red] (8.7,5.2) node[circle]{Section \#};
\draw [color=red,anchor=west] (8.7,2.8+0.1) node[circle]{3};
\draw [color=red,anchor=west] (8.7,0.45+0.1) node[circle]{2};
\draw [color=red,anchor=west] (8.7,-1.85+0.1) node[circle]{1};

\draw[arrows={latex-latex},color=white,line width=0.5pt] (5.4-0.1,3.325-0.025) -- (6.45-0.1,3.325-0.025);
\draw[arrows={latex-latex},color=white,line width=0.5pt] (5.45-0.125,2.3-0.025) -- (5.45-0.125,3.35-0.025);
\draw [color=white] (5.925-0.1,3.8) node[circle]{\Large $a$};
\draw [color=white] (5.0-0.1,2.875) node[circle]{\Large $a$};

\draw[arrows={latex-latex},color=white,line width=0.5pt] (-1.15,-1.32) -- (0.9,-1.32);
\draw [color=white] (0.9,-0.6) node[circle]{\Large $d_{\mathrm{cav}}$};

\end{tikzpicture}

\caption{QNM field distribution ($|E_y|$) in a cavity side-coupled to a W1 waveguide in a 2D rectangular PhC lattice with lattice constant $a$. Dashed lines in panel (a) separate periodic sections with distinct sets of Bloch modes, and Section 2 is the cavity section for constructing the roundtrip matrix, see Fig.~\ref{Fig:GeometryCavityRoundtrip}. The center-to-center distances between the cavities and the W1 waveguide are $d_{\mathrm{cav}} = 2a$ and $d_{\mathrm{cav}} = 3a$ in panels (a) and (b), respectively, and the associated QNM frequencies and $Q$ factors are given in Table~\ref{Tab:QNM_SideCouple}.}
\label{Fig:QNM_SideCouple}
\end{figure}
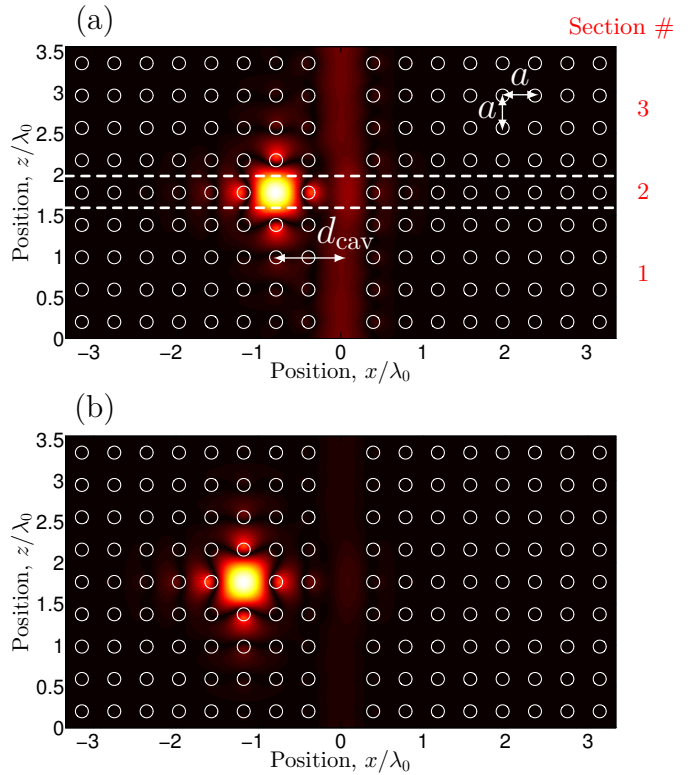

Removing a rod in the bulk of the PhC and in the vicinity of the W1 waveguide forms a side-coupled cavity, as illustrated in Fig.~\ref{Fig:QNM_SideCouple}; in the following we focus on determining QNMs of this structure. In panel~(a), dashed lines separate the periodic sections, each described by an underlying supercell (see Fig.~\ref{Fig:GeometryFMM}), for which distinct sets of Bloch modes (see Eqs.~\eqref{Eqn:FieldExpansions} and \eqref{Eqn:BlochModes}) are determined. Section 2 is the cavity section, for which a unity eigenvalue of the roundtrip matrix is determined.

By calculating the power reflection $R$ of the propagating Bloch mode in the W1 waveguide, indicated in the inset in Fig.~\ref{Fig:PhC_Reflection}, as function of a real frequency, we observe a peak, which we attribute to the excitation of a QNM. We use the value of the frequency at the reflection maximum as a starting point for the iteration towards a complex frequency giving a unity eigenvalue of the roundtrip matrix, see Eq.~\eqref{Eqn:RoundtripMatrix}. Specifically, we compute all eigenvalues of the roundtrip matrix $\vekn{M}(\omegat)$ and choose the eigenvalue that deviates the least from 1. We use this eigenvalue and a Newton-Raphson algorithm to iterate the complex frequency $\omegat$ until the eigenvalue of $\vekn{M}(\omegat)$ closest to unity deviates by less than a chosen tolerance, taken here as $\sim 10^{-12}$. For four different values of the cavity-W1 distance $d_{\mathrm{cav}}$ we determine QNMs, and the field distributions ($|E_y|$) of two of these are shown in Fig.~\ref{Fig:QNM_SideCouple}; their complex frequencies and $Q$ factors are given in Table~\ref{Tab:QNM_SideCouple}. 

\begin{table}
\centering
\begin{tabular}{c|c|c|c|c}
Fig.~\ref{Fig:QNM_SideCouple} & $d_{\mathrm{cav}}$ [$a$] & $\mathrm{Re}\left(\omegat\right)$ [$2\pi \mathrm{c} / a$] & $\mathrm{Im}\left(\omegat\right)$ [$2\pi \mathrm{c} / a$] & $Q$ \\
 \hline 
(a) & 2 & 0.397 & -0.0014 & $1.5 \cdot 10^2$ \\
(b) & 3 & 0.395 &  -0.00012 &  $1.7 \cdot 10^3$ \\
-- & 4 & 0.395 & -0.0000097 & $2.0 \cdot 10^4$ \\
-- & 5 & 0.395 & -0.00000077 & $2.5 \cdot 10^5$ \vspace{0.3cm}\\
\end{tabular}
\caption{QNM frequencies and $Q$ factors for cavities side-coupled to W1 waveguide structure at different cavity distances $d_{\mathrm{cav}}$. Two of the associated QNM field distributions are shown in Fig.~\ref{Fig:QNM_SideCouple}.} \label{Tab:QNM_SideCouple}
\end{table}

The real part of the QNM frequency $\mathrm{Re}(\omegat)$ remains essentially constant as $d_{\mathrm{cav}}$ is increased. In contrast, the absolute value of the imaginary part of the QNM frequency $\mathrm{Im}(\omegat)$ decreases by more than an order of magnitude when $d_{\mathrm{cav}}$ is increased by a lattice constant, giving rise to $Q$ factors that increase similarly as the cavity is moved away from the waveguide. This increase of the $Q$ factor reflects a smaller rate of leakage from the cavity into the waveguide as $d_{\mathrm{cav}}$ becomes larger. The 2D calculations presented here lack the out-of-plane ($y$) contribution to the corresponding 3D $Q$ factor, but the approach for determining QNMs and their $Q$ factors is readily extendable to 3D.

The type of structure considered in this section is dominated by a single QNM, and we may reconstruct the reflection spectrum in Fig.~\ref{Fig:PhC_Reflection} as a Lorentzian parametrized by the QNM-frequency
\begin{align} \label{Eqn:RQNM}
R(\omega) = \dfrac{\left[\mathrm{Im}\left(\omegat\right)\right]^2}{\left[\omega - \mathrm{Re}\left(\omegat\right)\right]^2 +\left[\mathrm{Im}\left(\omegat\right)\right]^2},
\end{align}
that is shown as the blue solid curve in Fig.~\ref{Fig:PhC_Reflection}. For frequencies within a linewidth of the peak ($R > 0.5$), the deviation between the scattering calculation and the QNM-approximated spectrum (Eq.~\eqref{Eqn:RQNM}) is less than 1\%, whereas the error increases further away from the cavity resonance frequency. Since the complex QNM-frequencies can typically be obtained with fewer computations than the full scattering spectrum, QNMs thus constitute a simple and practical way of obtaining the spectrum and the $Q$ factor of the resonator.

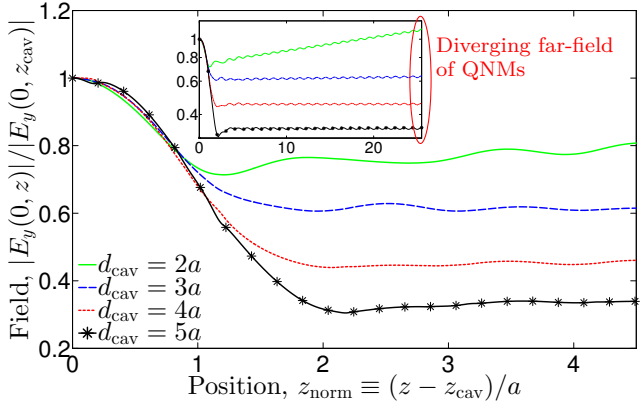
\begin{figure}[t!]
\centering 

\begin{tikzpicture} [line cap=round,line join=round,>=triangle 45,x=1.0cm,y=1.0cm, scale=0.47]
%


\node (img) at (0,0){\includegraphics[scale=0.22]{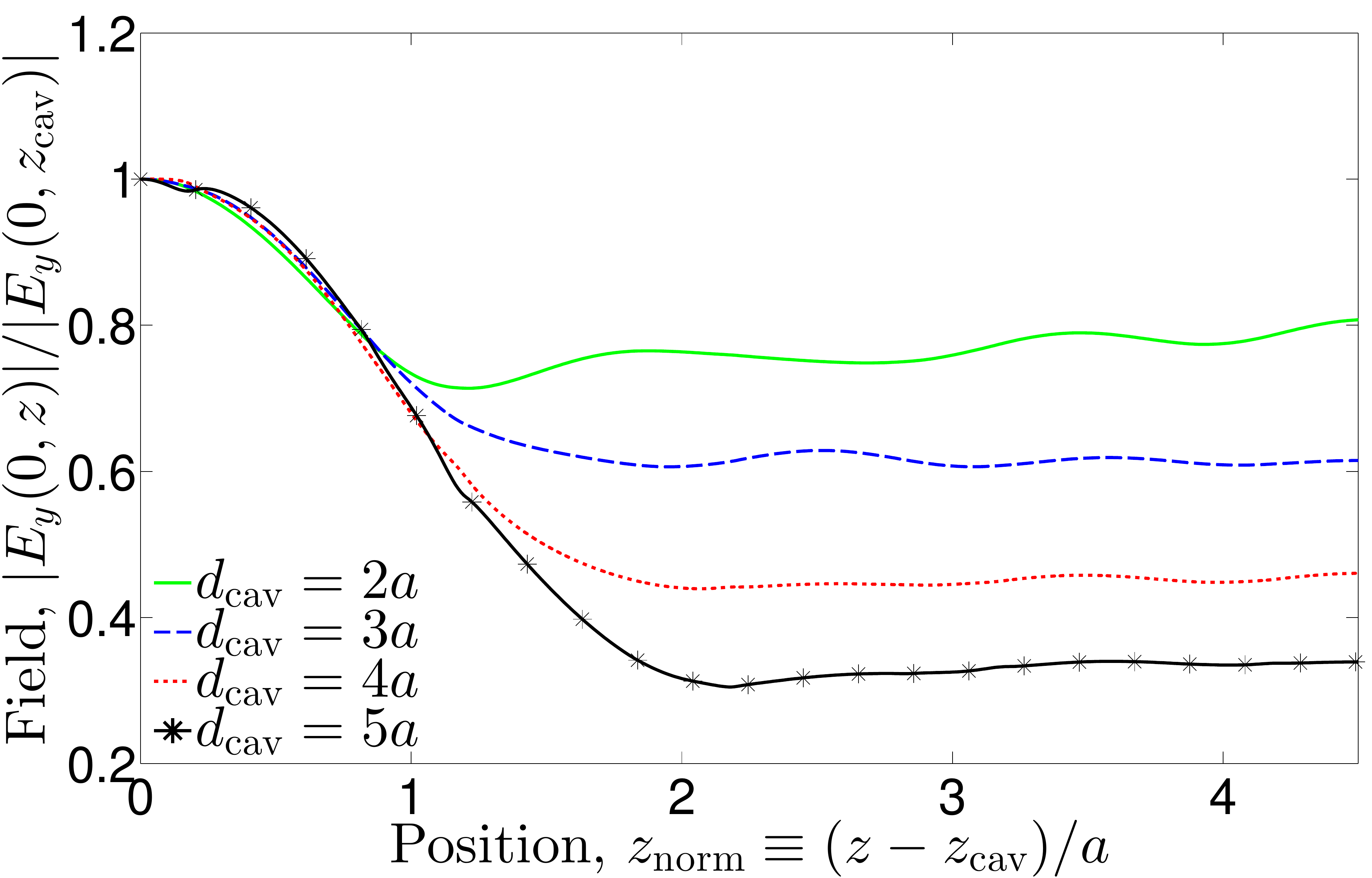}};
\node (img) at (-0.7,2.9+0.25){\includegraphics[scale=0.075]{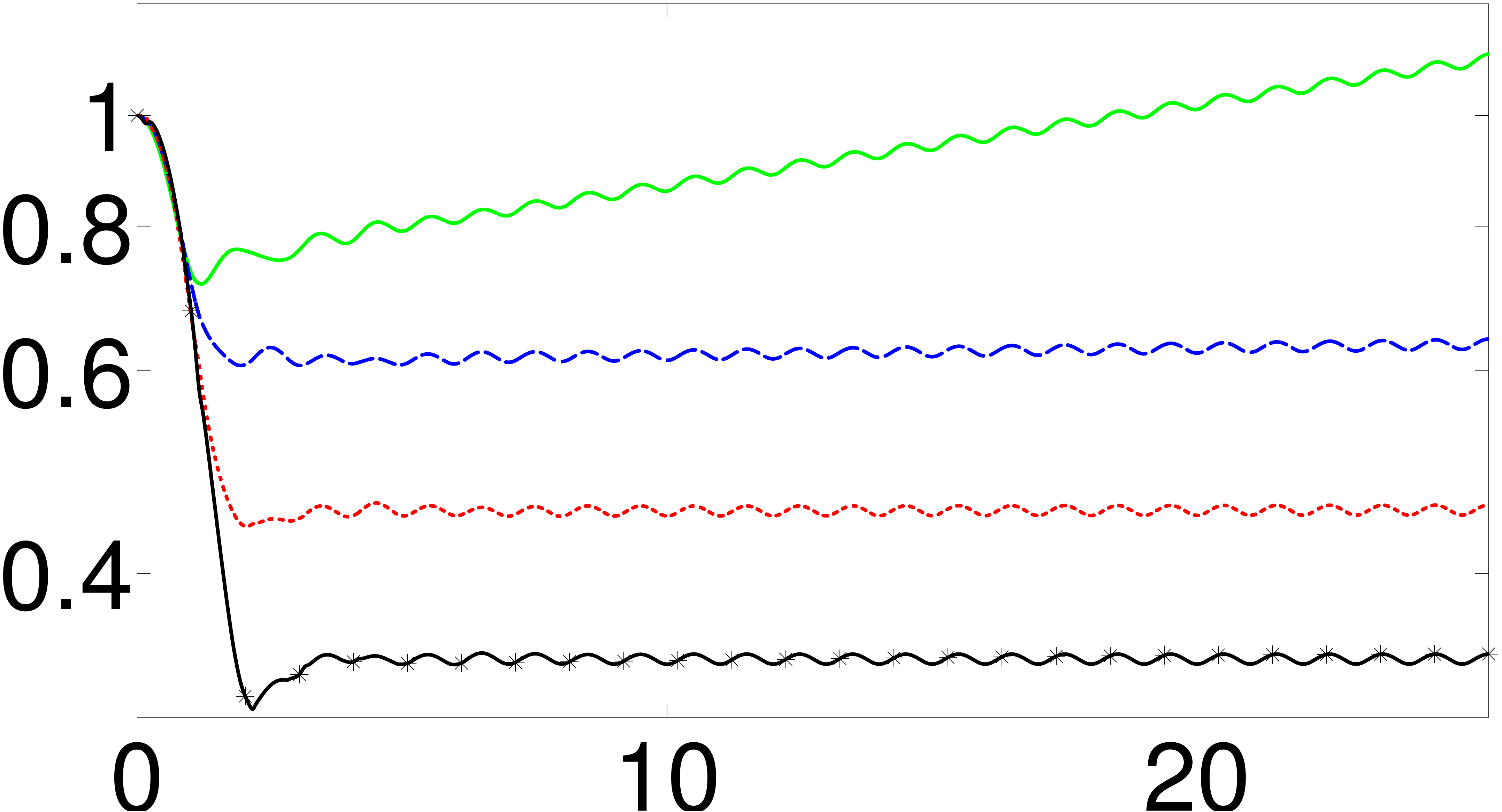}};

\draw [color=red] (2.7,3.2+0.15) ellipse (0.3cm and 1.8cm);

\node[anchor=west,color=red] at (2.9,4.1+0.2) (description) {\footnotesize Diverging far-field};
\node[anchor=west,color=red] at (2.9,3.4+0.2) (description) {\footnotesize of QNMs};

\end{tikzpicture}

\caption{Normalized QNM field distribution in a cavity side-coupled to a W1 waveguide in a 2D rectangular PhC lattice in the middle of the W1 waveguide ($x = 0$) as function of $z_{\mathrm{norm}} \equiv (z-z_{\mathrm{cav}})/a$. Different curves correspond to different cavity-W1 distances $d_{\mathrm{cav}}$ (see Fig.~\ref{Fig:QNM_SideCouple}(a)). The figure shows the QNMs in the near-field, corresponding to the $z$ coordinates used in Fig.~\ref{Fig:QNM_SideCouple}, while the inset includes the far-field behavior of the QNMs in a semilogarithmic plot.}
\label{Fig:QNMvsZ_SideCouple}
\end{figure}

In Fig.~\ref{Fig:QNMvsZ_SideCouple}, we plot the normalized QNM field distribution in the middle of the waveguide ($x=0$) as function of $z_{\mathrm{norm}} \equiv (z-z_{\mathrm{cav}})/a$, where $z_{\mathrm{cav}}$ is the cavity $z$ coordinate. The figure shows the near-field behavior, corresponding to the $z$ coordinates used in Fig.~\ref{Fig:QNM_SideCouple}, while the inset includes the far-field behavior of the QNMs in a semilogarithmic plot. In the near-field, the modes exhibit a maximum at $z_{\mathrm{norm}} = 0$ for all values of $d_{\mathrm{cav}}$, and the periodic modulation due to the Bloch mode form in Eqs.~\eqref{Eqn:BlochModes} is clearly visible. In the far-field, the periodic modulation of the QNMs is preserved due to the infinitely extended PhC waveguide, but more interestingly the exponential divergence of the envelope of the QNMs is obvious, in particular for the QNM with $d_{\mathrm{cav}} = 2a$ (green solid curve). The envelope of the QNM with $d_{\mathrm{cav}} = 3a$ (blue dashed curve) increases slightly, while for the largest-$Q$ QNMs ($d_{\mathrm{cav}} = 4a$, red dotted curve, and $d_{\mathrm{cav}} = 5a$, black solid-star curve), the divergence is not visible at these distances. Importantly, the magnitude of neither of the fields tends to zero in the far-field.

\subsection{Cavity in-line-coupled to W1 waveguide} \label{Sec:PhCInlineQNM}
\begin{figure}[t!]
\centering 

\begin{tikzpicture} [line cap=round,line join=round,>=triangle 45,x=1.0cm,y=1.0cm, scale=0.47]
%


\node (img) at (0,0){\includegraphics[scale=0.19]{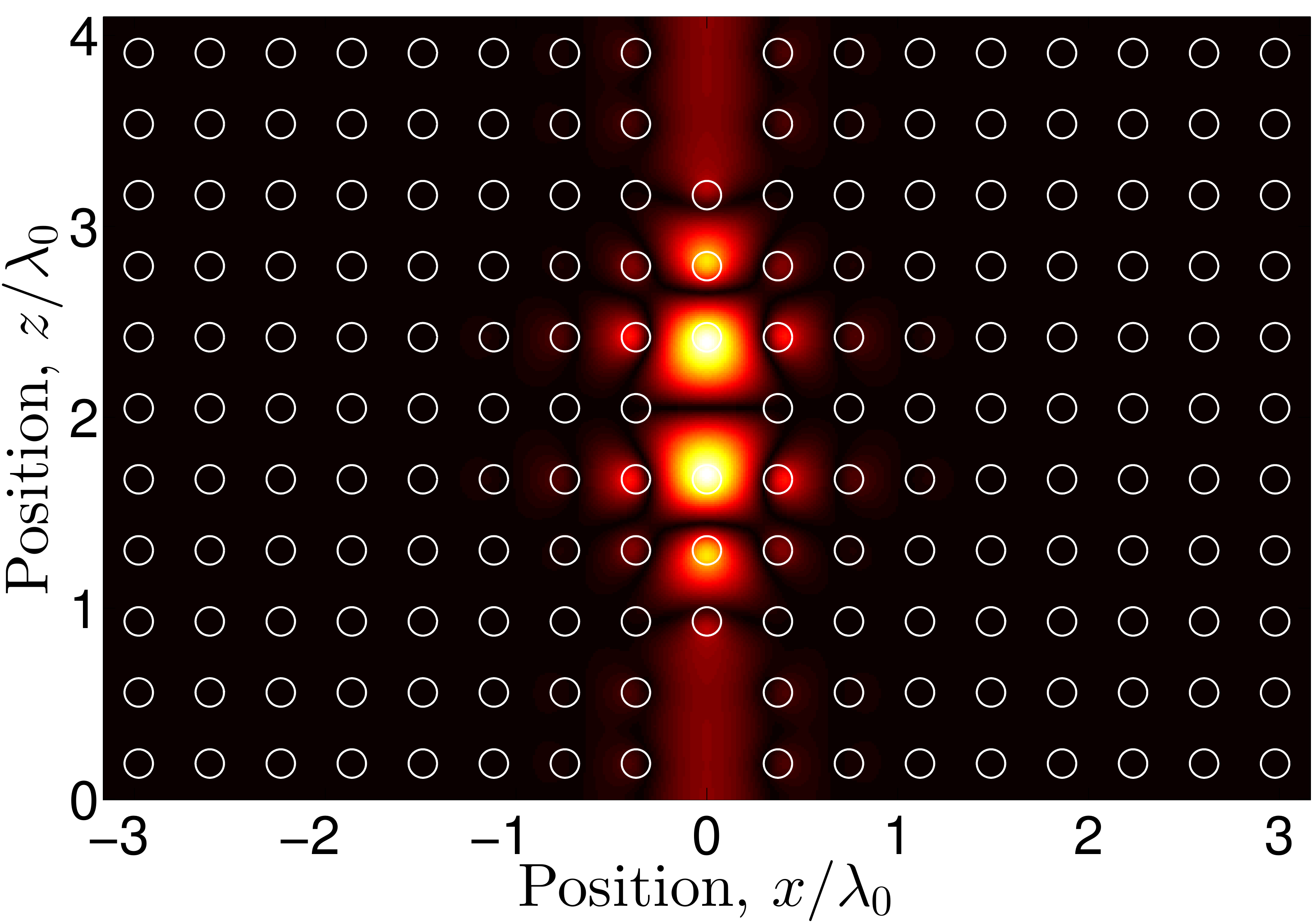}};

\draw[-,color=white,line width=1pt,dashed] (-7.38+1.525,-2.2+0.1) -- (8.55-1.65,-2.2+0.1);

\draw[-,color=white,line width=1pt,dashed] (-7.38+1.525,-1.40+0.1) -- (8.55-1.65,-1.40+0.1);

\draw[-,color=white,line width=1pt,dashed] (-7.38+1.525,-0.65+0.1) -- (8.55-1.65,-0.65+0.1);

\draw[-,color=white,line width=1pt,dashed] (-7.38+1.525,0.1+0.1) -- (8.55-1.65,0.1+0.1);

\draw[-,color=white,line width=1pt,dashed] (-7.38+1.525,0.85+0.1) -- (8.55-1.65,0.85+0.1);

\draw[-,color=white,line width=1pt,dashed] (-7.38+1.525,1.6+0.1) -- (8.55-1.65,1.6+0.1);

\draw[-,color=white,line width=1pt,dashed] (-7.38+1.525,2.35+0.1) -- (8.55-1.65,2.35+0.1);

\draw[-,color=white,line width=1pt,dashed] (-7.38+1.525,3.15) -- (8.55-1.65,3.15);

\draw [color=red] (8.7-1.65,5.2) node[circle]{Section \#};
\draw [color=red,anchor=west] (8.7-1.65,-2.9+0.1) node[circle]{1};
\draw [color=red,anchor=west] (8.7-1.65,-1.75+0.1) node[circle]{2};
\draw [color=red,anchor=west] (8.7-1.65,-1+0.1) node[circle]{3};
\draw [color=red,anchor=west] (8.7-1.65,-0.25+0.1) node[circle]{4};
\draw [color=red,anchor=west] (8.7-1.65,0.5+0.1) node[circle]{5};
\draw [color=red,anchor=west] (8.7-1.65,1.25+0.1) node[circle]{6};
\draw [color=red,anchor=west] (8.7-1.65,2+0.1) node[circle]{7};
\draw [color=red,anchor=west] (8.7-1.65,2.75+0.1) node[circle]{8};
\draw [color=red,anchor=west] (8.7-1.65,3.9+0.1) node[circle]{9};

\end{tikzpicture}

\caption{QNM field distribution ($|E_y|$) in a cavity in-line-coupled to a W1 waveguide in a 2D rectangular PhC lattice. The cavity is formed by surrounding one row (section 5) by blocking elements in the waveguide, constituted by sections 2, 3 and 4 (bottom mirror) and sections 6, 7 and 8 (top mirror). The refractive index of the blocking elements is varied linearly as specified in Eqs.~\eqref{Eqn:IndexAdiabaticMirror}. The QNM is independent of the choice of cavity section, cf. Table~\ref{Tab:QNM_InLineCouple}.}
\label{Fig:QNM_InLineCouple}
\end{figure}
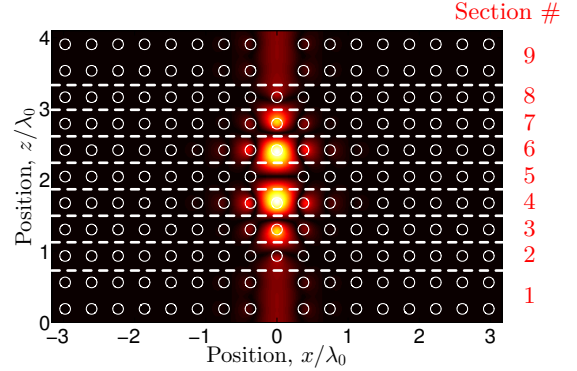

As a second example, we consider again the 2D rectangular PhC lattice with a W1 waveguide. We remove the side-coupled cavity considered in the previous section and instead implement an in-line cavity. This is done by surrounding a single row of the W1 waveguide by mirrors constituted of blocking elements, as shown in Fig.~\ref{Fig:QNM_InLineCouple}. Sections 1 and 9 are the waveguide sections in which the QNM is outgoing and diverging, and sections 2, 3 and 4 (6, 7 and 8) constitute the bottom (top) mirror surrounding the central cavity section 5. Additionally, we vary the refractive indices of the blocking elements in the waveguide in sections 2, 3, 4, 6, 7 and 8 linearly as 
\begin{subequations} \label{Eqn:IndexAdiabaticMirror}
\begin{align}
n_w &= n_{\mathrm{Back}} + \Delta_w \left( n_{\mathrm{Rods}} - n_{\mathrm{Back}} \right), \\
\Delta_2 &= \Delta_8 = 0.9, \,\,\, \Delta_3 = \Delta_7 = 0.6, \,\,\, \Delta_4 = \Delta_6 = 0.3.
\end{align}
\end{subequations}
with $n_{\mathrm{Rods}} = \sqrt{\epsilon_{\mathrm{Rods}}}$ and $n_{\mathrm{Back}} = \sqrt{\epsilon_{\mathrm{Back}}}$.

A QNM, with complex frequency $\omegat a / 2 \pi c = 0.375 - 0.0012 \ci$, may be readily calculated using Section 5 as the cavity section. However, as is apparent from the field distribution shown in Fig.~\ref{Fig:QNM_InLineCouple}, the mode leaks into the surrounding mirrors, and we may use other sections as the cavity in the roundtrip matrix method for determining the QNM. We use sections 2, 3 and 4 as the cavity and determine the QNM of the structure in Fig.~\ref{Fig:QNM_InLineCouple}. 
\begin{table}
\centering
\begin{tabular}{c|c|c|}
Cavity $w$ & $|\omegat_5 - \omegat_w|/|\omegat_5|$ & $\int | E_y^5 - E_y^w | \ddd \vekr/\int | E_y^5 | \ddd \vekr $ \\
 \hline 
5 &  0 &  0 \\
4 &  $7.8 \cdot 10^{-14}$ &   $1.6 \cdot 10^{-10}$ \\
3 &  $3.8 \cdot 10^{-14}$ &  $1.3 \cdot 10^{-10}$ \\
2 &  $2.4 \cdot 10^{-14}$ & $3.6 \cdot 10^{-10}$ \vspace{0.3cm}\\
\end{tabular}
\caption{Relative deviations of the QNM frequencies and near-field distributions for different choices of the cavity section $w$ for the in-line-coupled PhC cavity in Fig.~\ref{Fig:QNM_InLineCouple}. The QNM frequency is $\omegat a / (2\pi c) = 0.375 - 0.0012 \ci$.} \label{Tab:QNM_InLineCouple}
\end{table}
In Table~\ref{Tab:QNM_InLineCouple}, we give the corresponding relative deviations of the QNM frequencies and the near-field distributions wrt. the values obtained when using section 5 as the cavity. Both quantities are orders of magnitude smaller than unity, illustrating that different choices of the cavity section lead to the same QNM. This demonstrates the insensitivity to the choice of cavity in the roundtrip matrix method, which is an important property for structures that do not contain a well-defined cavity, e.g. the adiabatic micropillar cavity structure of~\cite{Lermer2012}.

\section{Conclusion}\label{Sec:Conclusion}
Quasi-normal modes are solutions of the source-free Maxwell equations satisfying an outgoing wave boundary condition (or radiation condition), and provide a natural framework for modeling of most resonant nanophotonic structures. The modes are characterized by complex frequencies, with the imaginary part accounting for their leaky nature; a property which translates into a spatial divergence when considering the propagating nature of the modes. We have presented a method for calculating quasi-normal modes of open nanophotonic structures using an internal cavity section and the associated cavity roundtrip matrix. The method is based on the use of scattering matrices, and we have developed the details of the method using Bloch mode expansions of the electromagnetic fields and the associated Bloch mode scattering matrices. We emphasize that other representations of the fields can be employed, for example Green's function expansions~\cite{Botten2000}, and that the importance lies in the access to scattering matrices in the chosen representation. As compared to previously developed methods relying on an inversion of the total scattering matrix, the method that we have presented here is more intuitive, being analogous to the well-known lasing condition, and easier to implement numerically.

We have demonstrated the use of the method by determining quasi-normal modes in two types of two-dimensional photonic crystal cavities coupled to infinitely extended waveguides. The use of Bloch mode expansions allows the modeling of infinitely extended structures without imposing absorbing boundary conditions like perfectly matched layers. To the best of our knowledge, explicit calculation of quasi-normal modes in such infinitely periodic structures have not been demonstrated before. For the second example of quasi-normal modes in photonic crystal cavities, we also demonstrated the robustness of the method wrt. the choice of the internal cavity section. This is particularly important when employing the method for structures that do not feature a well-defined cavity, as for example the adiabatic micropillar cavity structure of~\cite{Lermer2012}.

As an outlook, the quasi-normal modes of infinitely extended photonic crystals may provide a rigorous basis for describing light propagation in these structures, which may be of importance for understanding optical switching~\cite{Nozaki2010,Rephaeli2012,Heuck2013a,Fan2002,Yu2013}. Also, by including a finite imaginary part of the permittivity, structures with material gain (or loss) can be analyzed. It was recently shown~\cite{Grgic2012} that such structures display a strong coupling between gain and dispersion, which the method described here is ideally suited for analyzing further. We recall the theories based on quasi-normal modes for modeling of dipole sources~\cite{Kristensen2012,Sauvan2013,Ge2013}, for modeling of one-dimensional photonic crystals~\cite{Settimi2003,Settimi2009}, and for modeling of the scattering properties of structures of finite extent~\cite{Bai2013}, and envision that similar theories could be developed for coupled waveguide-cavity photonic crystal structures~\cite{Heuck2012,Kristensen2013}; this and normalization of QNMs in these structures is work in progress, which will appear in future publications. Such theories are particularly interesting for three-dimensional systems, for which full-scale optical simulations are known to be computationally demanding and time-consuming.

\appendix
\section{Determination and classification of Bloch modes} \label{App:ClassBlochModes}
To determine Bloch modes in the periodic sections, defined in Fig.~\ref{Fig:GeometryFMM}, we expand each Bloch mode on the lateral eigenmodes of the first layer in the supercell (left panel of Fig.~\ref{Fig:GeometryFMM}) and convert the Bloch mode condition to a generalized eigenvalue problem for the expansion coefficients and the Bloch mode eigenvalues $\rho_{j}^{w}$~\cite{Cao2002}. This produces the set of Bloch modes $\Omega = \lbrace \vekn{e}_j^{w}, \vekn{h}_j^{w} ; \rho_j^{w} \rbrace_j$ that we partition into waves in the upward ($+z$) $\Omega^{+} = \lbrace \vekn{e}_j^{w+}, \vekn{h}_j^{w+} ; \rho_j^{w+} \rbrace_j$ and in the downward ($-z$) $\Omega^{-} = \lbrace \vekn{e}_j^{w-}, \vekn{h}_j^{w-} ; \rho_j^{w-} \rbrace_j$ directions. The sets $\Omega^{+}$ and $\Omega^{-}$ contain the same number of modes, and for mirror symmetric supercells the modes in the two sets come in pairs, whose Bloch eigenvalues satisfy $\rho_j^{w+} = 1/\rho_j^{w-}$~\cite{Botten2001}. In general the Bloch modes can be either decaying, propagating or growing along the $z$ direction, and from Eq.~\eqref{Eqn:BlochModeRho} we note that propagating Bloch modes have purely real wave numbers $k_{j}^{w}$ while decaying and growing modes have complex or imaginary wave numbers. We provide an overview of the different types of Bloch modes in Table~\ref{Tab:Rho}.

\begin{table}
\begin{tabular}{c|c|c}
Mode Type & Upward ($+z$) & Downward ($-z$) \\
 \hline 
Decaying & $|\rho_j^{w+}| < 1$ & $|\rho_j^{w-}| > 1$ \\
Propagating & $|\rho_j^{w+}| = 1$ &  $|\rho_j^{w-}| = 1$ \\
Growing & $|\rho_j^{w+}| > 1$ & $|\rho_j^{w-}| < 1$ 
\end{tabular}
\caption{Overview of Bloch mode eigenvalues $|\rho_j^w|$ (see Eqs.~\eqref{Eqn:BlochModes}) in upward ($+z$) and downward ($-z$) direction. Bloch modes in both sets are classified as either decaying, propagating or growing.} \label{Tab:Rho}
\end{table}

We do not a priori know which Bloch modes belong to which of the two sets $\Omega^{+}$ and $\Omega^{-}$, but need to impose physical requirements to make this partitioning. To this end, we introduce the laterally integrated $z$ component of the Poynting vector of a Bloch mode, giving the net power $P_j^w$ in the propagation direction
\begin{align} \label{Eqn:Power}
P_j^w \equiv \dfrac{1}{2}  \int \mathrm{Re} \bigg [ \vekn{e}_j^w \times \left(\vekn{h}_j^w\right)^* \bigg ] \cdot \vekn{\hat{z}} \ddd \vekr_\perp.
\end{align}
With $P_j^w > 0$ ($P_j^w < 0$) the net power flow of the mode is in the $+z$ ($-z$) direction. We refer to the set of decaying and the set of propagating or growing Bloch modes as $\Omega_{\mathrm{D}} = \Omega_{\mathrm{D}}^{+} + \Omega_{\mathrm{D}}^{-}$ and $\Omega_{\mathrm{PG}} = \Omega_{\mathrm{PG}}^{+} + \Omega_{\mathrm{PG}}^{-}$, respectively. To partition the Bloch modes automatically in different situations, we have developed the following classification algorithm, which is a generalization of that proposed in~\cite{Botten2001}:\vspace{0.2cm}

\begin{enumerate}
\item Classify decaying modes ($\Omega_{\mathrm{D}}$): \label{Point:Decaying}
\begin{enumerate}
\item Upward ($+z$), $\Omega_{\mathrm{D}}^{+}$: $|\rho_j^w|<1$ $\land$ $\left| 1/ |\rho_j^w| - 1 \right| > \delta$ \label{Point:UpwardDecaying}
\item Downward ($-z$), $\Omega_{\mathrm{D}}^{-}$: $|\rho_j^w|>1$ $\land$ $\left||\rho_j^w| - 1 \right| > \delta$
\end{enumerate}
\item Classify propagating and growing modes ($\Omega_{\mathrm{PG}}$): \label{Point:Prop}
\begin{enumerate}
\item Upward ($+z$), $\Omega_{\mathrm{PG}}^{+}$: $\Omega \, \backslash \, \Omega_{\mathrm{D}}$ $\land$ $P_{j}^{w} > 0$ \label{Point:UpwardProp}
\item Downward ($-z$), $\Omega_{\mathrm{PG}}^{-}$: $\Omega \backslash \Omega_{\mathrm{D}}$ $\land$ $P_{j}^{w} < 0$
\end{enumerate}
\end{enumerate}
In the simplest case of a passive structure, at a real frequency and without PMLs, the propagating Bloch modes would ideally have $| \rho_{j}^{w} | = 1$, but due to numerical rounding errors these deviate slightly from unity. To account for this we introduce the empirical parameter $\delta$, with $0 < \delta \ll 1$ and $\delta$ typically on the order of $10^{-3}$ or smaller. 

When we change the simple structure, for instance by introducing PMLs, by adding loss or gain or by adding an imaginary part to the frequency, we perturb the set of Bloch modes, and in particular $| \rho_{j}^{w} |$ of the propagating Bloch modes start to deviate from unity by more than numerical rounding errors. Thus, we increase $\delta$ to include these quasi-propagating Bloch modes in the power sorting, cf. point~\ref{Point:Prop} in the above sorting algorithm. Since the difference in $| \rho_{j}^{w} |$ between propagating and decaying Bloch modes in the simple situation may be very small, increasing $\delta$ in the perturbed situations might result in inclusion of decaying Bloch modes in the wrong set~\cite{PowerDecayingModes}. In the PhCs analyzed in Section~\ref{Sec:QNMsPhCs}, the difference in $|\rho_{j}^{w}|$ between propagating and decaying Bloch modes is sufficiently large that the classification of Bloch modes is unambiguous. In other structures, the difference in $| \rho_{j}^{w} |$ between propagating and decaying Bloch modes may be a lot smaller, and in such cases it is necessary to determine the QNMs with different values of $\delta$ to ensure that the Bloch mode sorting does not influence the QNM complex frequency and field distribution. Also, the lower the $Q$ factor of the QNM to be determined is, the higher is the value of $\delta$ needed to ensure convergence of the iterative routine to find the complex QNM frequency.

\section{Equivalence of total scattering matrix and cavity roundtrip matrix methods in one-dimensional three-section structure} \label{App:EquivalenceMethods}
In the general case, we cannot show analytically that the total scattering matrix method (Section~\ref{Sec:ScatteringMethod}) and the cavity roundtrip matrix method (Section~\ref{Sec:RoundtripMethod}) lead to the same complex QNM frequencies since we cannot write simple, closed-form expressions for the involved matrices. However, for certain simple structures we can write compact expressions for these matrices, and in this appendix we explore one such case and show that the two methods lead to the same equation for the complex QNM frequency. 

We consider a structure consisting of three laterally uniform and $z$ invariant sections in which the Bloch modes are the lateral eigenmodes that in turn are plane waves. We assume normal incidence which makes it a 1D problem. At the interfaces, the reflection and transmission of each of the plane waves are determined by the Fresnel coefficients~\cite{Jackson1998Chap7}, and using section 2 as the cavity section, cf. Fig.~\ref{Fig:GeometryCavityRoundtrip}, the requirement of a unity eigenvalue of the roundtrip matrix, introduced in Eq.~\eqref{Eqn:RoundtripMatrix}, becomes
\begin{align} \label{Eqn:RoundtripThreeLayer}
M(\omegat) = R^{2,1} P^2 R^{2,3} P^2 = 1.
\end{align}
Here, $P^2 = \exp \left(\ci n_2 k_0 h_2 \right)$ where $n_2$ and $h_2$ are the refractive index and the height of section 2, respectively. $R^{2,1}$ and $R^{2,3}$ are Fresnel reflection coefficients. 

The elements of the total scattering matrix, that relate the outgoing to the incoming amplitudes in sections 1 and 3
\begin{subequations}
\begin{align}
\begin{bmatrix} c^{1,-} \\ c^{3,+} \end{bmatrix} &= 
\begin{bmatrix}
S^{1,1} & S^{1,2} \\
S^{2,1} & S^{2,2}
\end{bmatrix}
\begin{bmatrix} c^{1,+} \\ c^{3,-} \end{bmatrix},
\end{align}
read as follows~\cite{Lavrinenko2014Chap6}
\begin{align}
S^{1,1} &= R^{1,2} + T^{2,1} P^2 R^{2,3} P^2 \left(1 - R^{2,1} P^2 R^{2,3} P^2 \right)^{-1} T^{1,2}, \\
S^{1,2} &= T^{2,1} P^2 T^{3,2} + T^{2,1} P^2 R^{2,3} P^2 \nonumber \\
&\hspace{0.65cm}\times \left(1 - R^{2,1} P^2 R^{2,3} P^2 \right)^{-1} R^{2,1} P^2 T^{3,2}, \\
S^{2,1} &= T^{2,3} P^2 \left(1 - R^{2,1} P^2 R^{2,3} P^2 \right)^{-1} T^{1,2}, \\
S^{2,2} &= R^{3,2} + T^{2,3} P^2 \left(1 - R^{2,1} P^2 R^{2,3} P^2 \right)^{-1} R^{2,1} P^2 T^{3,2}.
\end{align}
\end{subequations}
All $R$  and $T$ coefficients are Fresnel reflection or transmission coefficients. Provided the inverse scattering matrix exists, its determinant can be expressed as the inverse of the determinant of the scattering matrix
\begin{align}
\mathrm{det}\left( \vekn{S}^{-1} (\omegat) \right) &= \dfrac{1}{\mathrm{det}\left( \vekn{S} (\omegat) \right)} \nonumber \\ 
&= \dfrac{\left[\left(1-R^{2,1} P^2 R^{2,3} P^2\right)\right]^2}{(\dots )} ,
\end{align}
where, in the view of Eq.~\eqref{Eqn:QNMConditionLit2}, the detailed expression for the denominator is left out; importantly, it does not vanish when $R^{2,1} P^2 R^{2,3} P^2 \rightarrow 0$. Finally, using Eq.~\eqref{Eqn:QNMConditionLit2} the total scattering matrix method equation for the complex QNM frequency reads
\begin{align}
R^{2,1} P^2 R^{2,3} P^2 = 1,
\end{align}
which is the same equation as obtained from the roundtrip matrix method in Eq.~\eqref{Eqn:RoundtripThreeLayer}. We have thus, in the simple case of three uniform sections, demonstrated that the two methods lead to the same equation for the complex QNM frequencies.

\section*{Acknowledgements}
This work was funded by project SIQUTE (contract EXL02) of the European Metrology Research Programme (EMRP). The EMRP is jointly funded by the EMRP participating countries within EURAMET and the European Union. Support from the Villum Foundation via the VKR Centre of Excellence NATEC and the Danish Council
for Independent Research (FTP 10-093651) is gratefully acknowledged.


\end{document}